\documentclass[%
aps,
prx,
reprint,
superscriptaddress,
nofootinbib,
amsmath,
amssymb
]{revtex4-2}
\usepackage{
    physics,
    graphicx,
    multirow,
    tabularx,
    mathtools, %for \begin{rcases} \end{rcases}
    placeins, %To include \FloatBarrier
    nicefrac, %For inline fraction
    hyperref %To use put in url, enable links in bibiliography
}

\usepackage[capitalize]{cleveref}
\Crefname{section}{Sec.}{Secs.}

\usepackage[dvipsnames]{xcolor}
\hypersetup{
    colorlinks,
    linkcolor={blue!50!black},
    citecolor={blue!50!black},
    urlcolor={blue!80!black}
}% remove the box around citation link and bibilio link

\usepackage[caption=false]{subfig} %The caption part is not compatible with revtex
\newcommand{\fiteqn}[1]{\resizebox{\hsize}{!}{$#1$}} %Used within math env to fit single line eqn
\renewcommand{\var}[1]{{\mathrm{Var}[#1]}}
\newcommand{\cov}[1]{{\mathrm{Cov}[#1]}}
\newcommand{\mse}[1]{{\mathrm{MSE}[#1]}}
\newcommand{\bias}[1]{{\mathrm{Bias}[#1]}}
\newcommand{\expect}[1]{{\mathrm{E}[#1]}}
\newcommand{\prob}[1]{{\mathrm{P}[#1]}}
\newcommand{\range}[1]{{\mathrm{R}[#1]}}
\newcommand{\mean}[1]{{\overline{#1}}}
\newcommand{\est}[1]{{\hat{#1}}}

\DeclarePairedDelimiterX\pbraket[2]{\langle\!\langle}{\rangle\!\rangle}{#1 \delimsize\vert #2}
\DeclareMathOperator*{\sign}{sgn}

\begin{document}
    
%\preprint{APS/123-QED}

\title{A Practical Framework for Quantum Error Mitigation}% Force line breaks with \\

%\author{Authors}
%\affiliation{Affiliations}

\author{Zhenyu Cai}
\email{cai.zhenyu.physics@gmail.com}
\affiliation{Department of Materials, University of Oxford, Oxford, OX1 3PH, United Kingdom}
\affiliation{St John's College, University of Oxford, Oxford, OX1 3JP, United Kingdom}
\affiliation{Quantum Motion Technologies Ltd, Nexus, Discovery Way, Leeds, LS2 3AA, United Kingdom}

\date{\today}% It is always \today, today,

\begin{abstract}
    Quantum error mitigation is expected to play a crucial role in the practical applications of quantum machines for the foreseeable future. Thus it is important to put the numerous quantum error mitigation schemes proposed under a coherent framework that can highlight their underlying connections while providing guidance for their practical performance. In this article, we construct a general framework named \emph{linear quantum error mitigation} that includes most of the state-of-the-art quantum error mitigation schemes. Within the framework, quantum error mitigation can be effectively viewed as extracting the error-mitigated state out of the noisy state, which introduces a new metric called \emph{extraction rate} for indicating the cost-effectiveness of a given mitigation scheme. Using the framework, we have derived and compared the extraction rate, improvement in the fidelity and sampling overhead across various mitigation schemes under practical assumptions. The structure, insights and intuitions provided by the framework can serve as a basis for the further developments of new schemes. 
\end{abstract}

\maketitle

%\tableofcontents
\section{Introduction}\label{sec:intro}

Noise in quantum computing devices has always been the key bottleneck for their practical implementation. Quantum error correction (QEC)~\cite{terhalQuantumErrorCorrection2015} paves the way towards fault-tolerant computation, however the qubit overhead required for their full implementation might still be decades away. Quantum error mitigation (QEM) provides a near-term alternative, in which we will probe the effects of errors on our target circuit and use post-processing to mitigate their damages~\cite{endoHybridQuantumClassicalAlgorithms2021}. Quantum error mitigation has been successfully implemented in many experiments~\cite{kandalaErrorMitigationExtends2019,sagastizabalExperimentalErrorMitigation2019,songQuantumComputationUniversal2019,googleaiquantumandcollaboratorsHartreeFockSuperconductingQubit2020,googleaiquantumandcollaboratorsObservationSeparatedDynamics2020}, and is expected to be a key component for practical quantum applications from the noisy intermediate-scale quantum (NISQ) era that we are entering now, all the way up to the moment right before we have full QEC fault-tolerance~\cite{suzukiQuantumErrorMitigation2021,piveteauErrorMitigationUniversal2021} in which error is no longer a bottleneck. 

Numerous QEM schemes have been proposed in recent years with examples like probabilistic error cancellation~\cite{temmeErrorMitigationShortDepth2017,endoPracticalQuantumError2018}, zero-noise extrapolation~\cite{liEfficientVariationalQuantum2017,temmeErrorMitigationShortDepth2017,endoPracticalQuantumError2018,caiMultiexponentialErrorExtrapolation2021}, subspace expansion~\cite{mccleanHybridQuantumclassicalHierarchy2017}, symmetry verification~\cite{mcardleErrorMitigatedDigitalQuantum2019,bonet-monroigLowcostErrorMitigation2018,caiQuantumErrorMitigation2021a} and purification-based methods~\cite{koczorExponentialErrorSuppression2021,hugginsVirtualDistillationQuantum2021,huoDualstatePurificationPractical2021,caiResourceefficientPurificationbasedQuantum2021}. However, these methods are mostly studied in a separate manner and it is hard to understand and compare them in a single coherent framework. 

More recently, there are attempts to combine some subsets of QEM techniques to better understand their connections~\cite{caiMultiexponentialErrorExtrapolation2021,bultriniUnifyingBenchmarkingStateoftheart2021,mariExtendingQuantumProbabilistic2021,loweUnifiedApproachDatadriven2021}. In particular, Ref.~\cite{wangCanErrorMitigation2021,takagiFundamentalLimitationsQuantum2021} attempt to derive general performance bounds for QEM, which is useful for understanding the fundamental limitations of QEM schemes. However, such bounds can be hard to calculate in practice for realistic noise models, and they might not be tight in practical settings as shown in numerical simulations~\cite{takagiFundamentalLimitationsQuantum2021}. Hence, such bounds might be too opaque to provide physical intuitions and insights about the practical performance of QEM.

In this article, we are going to introduce a framework called \emph{linear quantum error mitigation} which will encompass most of the state-of-the-art QEM methods. It will allow us to gain intuition about their underlying mechanism and give us the relevant metric to quantify their performance. In \cref{sec:qem_intro}, we are first going to introduce the general idea of QEM. Then we are going to introduce the framework of linear QEM in \cref{sec:qem_estimator}. By making assumptions about the state and the noise that are relevant to most practical settings, we can analytically derive the performance of different QEM methods in \cref{sec:gate_error_mitigation,sec:state_error_mitigation} and compare and combine them in \cref{sec:comp_comb}. We will end with a conclusion in \cref{sec:conclusion} and discuss the possible future directions.

\section{Quantum Error Mitigation}\label{sec:qem_intro}
We are given a circuit, called the \emph{primary circuit}, whose noiseless output has the expectation value $\Tr(O\rho_0)$, where $\rho_0$ is the ideal noiseless state from the primary circuit and $O$ is our observable of interests. In many practical applications, the ideal state $\rho_0$ is a \emph{pure state} and we will assume this is the case throughout this article, though many of our arguments later are also applicable to a mixed ideal state. Suppose we can estimate the target value $\Tr(O\rho_0)$ using some estimator $\est{O}$, then the expected deviation of the estimator from the target value is given by its \emph{mean square error}, which can be further decomposed into the \emph{bias} and the \emph{variance} of the estimator:
\begin{align*}
    \mse{\est{O}} &= \expect{(\est{O} - \Tr(O\rho_0))^2} = \bias{\est{O}}^2 + \var{\est{O}}
\end{align*}
with
\begin{align*}
    \bias{\est{O}} = \expect{\est{O}} - \Tr(O\rho_0),\quad \var{\est{O}} = \expect{\est{O}^2} - \expect{\est{O}}^2.
\end{align*}
In this article, when we say bias, we often refer to the absolute value of the bias. The meaning should be obvious from the context.

In practice, we will have a \emph{noisy primary circuit} of the noise level $\lambda$ (the exact definition is in \cref{sec:circ_fault}) and the corresponding noisy output state is denoted as $\rho_{\lambda}$. The random variable denoting the outcome of performing $O$ measurement on the noisy state $\rho_{\lambda}$ is denoted as $\est{O}_{\rho_{\lambda}}$ and the corresponding sample mean after $N_{\mathrm{cir}}$ circuit runs is denoted as $\mean{O}_{\rho_{\lambda}}$. If we naively use this noisy sample mean $\mean{O}_{\rho_{\lambda}}$ to estimate our target expectation value $\Tr(O\rho_0)$, then its mean square error is given by:
\begin{align*}
    \fiteqn{\mse{\mean{O}_{\rho_{\lambda}}} = \big(\underbrace{\Tr(O\rho_{\lambda}) - \Tr(O\rho_0)}_{\bias{\mean{O}_{\rho_{\lambda}}} = \bias{\est{O}_{\rho_{\lambda}}}}\big)^2 + \underbrace{\frac{\Tr(O^2\rho_{\lambda}) -  \Tr(O\rho_{\lambda})^2}{N_{\mathrm{cir}}}}_{\var{\mean{O}_{\rho_{\lambda}}}=\var{\est{O}_{\rho_{\lambda}}}/N_{\mathrm{cir}}}.}
\end{align*} 
The error contribution from the variance, which is called \emph{shot noise}, can be reduced by increasing the number of circuit runs (shots). Still, the bias limits the eventual precision of our results, even with an infinite number of circuit runs. 

With \emph{the same amount of circuit runs} $N_{\mathrm{cir}}$ allowed, we hope to reduce the bias in the estimator by constructing an error-mitigated estimator $\mean{O}_{\mathrm{em}}$:
\begin{align*}
    \bias{\mean{O}_{\mathrm{em}}} \leq \bias{\mean{O}_{\rho_{\lambda}}}.
\end{align*}
Unlike before in which we simply take the sample mean of the outcome of different circuit runs for $\mean{O}_{\rho_{\lambda}}$, now for $\mean{O}_{\mathrm{em}}$ we need to post-process the measurement outcome in a more complex way to extract and amplify the useful signal buried in the noise. Such amplifications will naturally increase the variance of the estimator:
\begin{align*}
    \var{\mean{O}_{\mathrm{em}} }\geq \var{\mean{O}_{\rho_{\lambda}}}. 
\end{align*}
We can define the effective `single-shot' error-mitigated estimator $\est{O}_{\mathrm{em}}$ whose mean with $N_{\mathrm{cir}}$ samples is simply $\mean{O}_{\mathrm{em}}$. Thus it will satisfy $\bias{\est{O}_{\mathrm{em}}} = \bias{\mean{O}_{\mathrm{em}}}$ and $\var{\est{O}_{\mathrm{em}}} = N_{\mathrm{cir}}\var{\mean{O}_{\mathrm{em}}}$. Hence, it will have a smaller bias and larger variance compared to the single-shot estimator directly from the noisy primary circuit:
\begin{align*}
    \bias{\est{O}_{\mathrm{em}}} &\leq \bias{\est{O}_{\rho_{\lambda}}},\quad\var{\est{O}_{\mathrm{em}} }\geq \var{\est{O}_{\rho_{\lambda}}}. 
\end{align*}

The sampling overhead of a given quantum error mitigation scheme is defined as:
\begin{align}\label{eqn:samp_cost}
    C_{\mathrm{em}} = \frac{\var{\mean{O}_{\mathrm{em}} }}{\var{\mean{O}_{\rho_{\lambda}}}} = \frac{\var{\est{O}_{\mathrm{em}} }}{\var{\est{O}_{\rho_{\lambda}}}}
\end{align}
which is simply the factor of increase in the number of circuit runs needed for the error-mitigated estimator to reach the same precision (shot noise level) as the unmitigated estimator.

The construction of an error-mitigated estimator can be largely decoupled into two parts:
\begin{itemize}
    \item \emph{Response measurement}: measures how the observable responds to the change of some given noise components in the circuit. 
    \item \emph{Noise calibration}: measures the strength of some given noise components in the circuit. 
\end{itemize}
Combining both steps will tell us how the noise in our primary circuit would damage our observable and thus enable us to construct an estimator with these damages removed.

\section{Linear Quantum Error Mitigation}\label{sec:qem_estimator}
We will now introduce the framework of \emph{linear quantum error mitigation}, which uses estimators that are linear in terms of the observable of interest $O$. It includes most of the mainstream QEM methods as we will discuss later.

We can always write a linear estimator $\est{O}_{\mathrm{em}}$ into the form
\begin{align*}
    \est{O}_{\mathrm{em}} = \Tr(O\est{\rho}_{\mathrm{em}})
\end{align*}
where $\est{\rho}_{\mathrm{em}}$ is the estimator for some operator $\rho_{\mathrm{em}}$. In most of the QEM methods, to ensure the normalisation of the estimator $\est{O}_{\mathrm{em}}$, $\rho_{\mathrm{em}}$ will have unit trace $\Tr(\rho_{\mathrm{em}}) = 1$ and is often (or very close to) a density operator. In these cases, QEM can be viewed as trying to extract an error-mitigated state $\rho_{\mathrm{em}}$ out of the original noisy state $\rho_{\lambda}$. 

A reduction in the bias of the error-mitigated estimator $\est{O}_{\mathrm{em}}$ over the unmitigated estimator $\est{O}_{\rho_{\lambda}}$ can usually be indicated by an increase in the fidelity (against the ideal state $\rho_0$) of the error-mitigated state $\rho_{\mathrm{em}}$ over the unmitigated state $\rho_{\lambda}$. In the special case that our observable is $O = \rho_0$, then the infidelities of $\rho_{\lambda}$ and $\rho_{\mathrm{em}}$ are exactly the biases of $\est{O}_{\rho_{\lambda}}$ and $\est{O}_{\mathrm{em}}$. Using fidelity measure as an indicator for estimation bias has also been proven to be effective via numerical simulations in some QEM schemes~\cite{hugginsVirtualDistillationQuantum2021,koczorExponentialErrorSuppression2021,caiQuantumErrorMitigation2021a}. Hence, here we will use the factor of fidelity boost going from the unmitigated state to the error-mitigated state as \emph{the indicator for the bias reduction} achieved by the QEM method:
\begin{align}\label{eqn:fid_boost_1}
    B_{\mathrm{em}} = \frac{\Tr(\rho_0\rho_{\mathrm{em}})}{\Tr(\rho_0\rho_{\lambda})}.
\end{align}
We will simply call $B_{\mathrm{em}}$ the \emph{fidelity boost}.

To see how we extract $\rho_{\mathrm{em}}$ out of $\rho_{\lambda}$, we can write the unmitigated noisy state $\rho_{\lambda}$ as a probabilistic mixture of the error-mitigated component $\rho_{\mathrm{em}}$ and the erroneous component $\rho_{\mathrm{err}}$:
\begin{align}\label{eqn:noisy_decomp}
    \rho_{\lambda} = p_{\mathrm{em}}\rho_{\mathrm{em}} + \left(1-p_{\mathrm{em}}\right)\rho_{\mathrm{err}},
\end{align}
where by definition we have $\Tr(\rho_{\mathrm{err}}) = 1$. Here $p_{\mathrm{em}}$ is chosen such that the erroneous component $\rho_{\mathrm{err}} = (\rho_{\lambda} - p_{\mathrm{em}}\rho_{\mathrm{em}})/(1-p_{\mathrm{em}})$ is orthogonal to the ideal state $\Tr(\rho_0\rho_{\mathrm{err}}) = 0$. Here $p_{\mathrm{em}}$ being the amount of the error-mitigated state contained in our noisy state will represent the amount of useful signal that we can extract. Taking the fidelity against $\rho_0$ for both sides of \cref{eqn:noisy_decomp} and compared to \cref{eqn:fid_boost_1}, we see that the fidelity boost is simply given by:
\begin{align}\label{eqn:fid_boost}
    B_{\mathrm{em}} = p_{\mathrm{em}}^{-1}= \frac{\Tr(\rho_0\rho_{\mathrm{em}})}{\Tr(\rho_0\rho_{\lambda})}.
\end{align}
Our definitions ensure that $B_{\mathrm{em}} \geq 1$ as long as $0\leq p_{\mathrm{em}} \leq 1$.

As mentioned, a key step in QEM is \emph{response measurement}. We can probe the response of the expectation value of the observable $O$ towards different error components by constructing different response measurement circuits by modifying the primary circuit. The output of the $i$th response measurement circuit can be written as $\Tr(O \sigma_{\mathrm{rsp}, i})$ where $\sigma_{\mathrm{rsp}, i}$ is the effective output `state' that is similar to the unmitigated state $\rho_{\lambda}$ but with a modified erroneous component. Note that $\sigma_{\mathrm{rsp}, i}$ is not necessarily a density operator, i.e. it may not be physical. The output `states' of the full set of response measurement circuits (including the primary circuit) is denoted as $\vec{\sigma}_{\mathrm{rsp}}$. If we take a probabilistic mixture of these circuits according to some probability distribution  $\vec{p}_{\mathrm{rsp}}$, then the effective output `state' is simply $\vec{p}_{\mathrm{rsp}}\cdot\vec{\sigma}_{\mathrm{rsp}}$. By choosing an appropriate distribution $\vec{p}_{\mathrm{rsp}}$, we can cancel out the erroneous components $\rho_{\mathrm{err}}$ in \cref{eqn:noisy_decomp} and extract \emph{part} of the error-mitigated components as the effective output state:
\begin{align}\label{eqn:em_extraction}
    q_{\mathrm{em}}\rho_{\mathrm{em}} = \vec{p}_{\mathrm{rsp}}\cdot\vec{\sigma}_{\mathrm{rsp}}
\end{align}
where $0\leq q_{\mathrm{em}} \leq p_{\mathrm{em}}$. The measurement outcome of this ensemble of response measurement circuits is denoted as $\est{O}_{\mathrm{rsp}}$ and its expectation value is simply
\begin{align}\label{eqn:rsp_cal}
    \expect{\est{O}_{\mathrm{rsp}}} = \Tr(O\vec{p}_{\mathrm{rsp}}\cdot\vec{\sigma}_{\mathrm{rsp}}) = \Tr(Oq_{\mathrm{em}}\rho_{\mathrm{em}}) = q_{\mathrm{em}} \expect{\est{O}_{\mathrm{em}}}.
\end{align}
Hence, the error-mitigated expectation value $\expect{\est{O}_{\mathrm{em}}}$ can be obtained from the error-mitigated estimator.
\begin{align}\label{eqn:linear_estimator}
    \est{O}_{\mathrm{em}} = q_{\mathrm{em}}^{-1}\est{O}_{\mathrm{rsp}}.
\end{align}
Correspondingly, we have:
\begin{align*}
    \var{\est{O}_{\mathrm{em}}} =  q_{\mathrm{em}}^{-2}\var{\est{O}_{\mathrm{rsp}}}.
\end{align*}
We would expect $\var{\est{O}_{\mathrm{rsp}}} \sim \var{\est{O}_{\rho_{\lambda}}}$ since the ensemble of calibration circuits that output $\est{O}_{\mathrm{rsp}}$ are variants of the primary circuit that output $\est{O}_{\rho_{\lambda}}$. Hence, we have:
\begin{align}\label{eqn:linear_var}
    \var{\est{O}_{\mathrm{em}}} \sim  q_{\mathrm{em}}^{-2}\var{\est{O}_{\rho_{\lambda}}}.
\end{align}
which corresponds to a sampling overhead of
\begin{align}\label{eqn:linear_samp_cost}
    C_{\mathrm{em}}  = \frac{\var{\est{O}_{\mathrm{em}}}}{\var{\est{O}_{\rho_{\lambda}}}} \sim q_{\mathrm{em}}^{-2} \geq p_{\mathrm{em}}^{-2} =  B_{\mathrm{em}}^{2}
\end{align}
using the definition in \cref{eqn:samp_cost}. We can also derive a similar result using Hoeffding's inequality as shown in \cref{sec:samp_hoeff}.

We can define the \emph{extraction rate} 
\begin{align}\label{eqn:extraction_rate}
    r_{\mathrm{em}} = \frac{q_{\mathrm{em}}}{p_{\mathrm{em}}} \sim \frac{B_{\mathrm{em}}}{\sqrt{C_{\mathrm{em}}}},
\end{align}
which is the fraction of error-mitigated state $\rho_{\mathrm{em}}$, out of all that contained in the unmitigated state $\rho_{\lambda}$, that has been successfully extracted via our QEM technique. It is also the fidelity boost we achieve per unit of the square root of the sampling overhead. Hence, the extraction rate can be used as an indicator of the `cost-effectiveness' of the QEM method.

In summary, for linear QEM methods, once we know the expression for the error-mitigated state $\rho_{\mathrm{em}}$, we can obtain its fidelity boost ($p_{\mathrm{em}}^{-1}$) using \cref{eqn:fid_boost} which serve as an indicator for the reduction in bias, and we can also obtain the lower bound on its sampling overhead ($p_{\mathrm{em}}^{-2}$) using \cref{eqn:linear_samp_cost}. The exact sampling overhead ($q_{\mathrm{em}}^{-2}$) can be obtained by studying the exact procedure we use to extract the error-mitigated state $\rho_{\mathrm{em}}$, which is simply \cref{eqn:em_extraction}. The cost-effectiveness of the QEM method can be indicated using the extraction rate in \cref{eqn:extraction_rate}. 

Instead of obtaining the error-mitigated result through the method above, if we have some way to post-select our circuit runs such that the effective output of the circuit is $\rho_{\mathrm{em}}$, then the sampling overhead of such method is simply the inverse of the post-selection rate $C_{\mathrm{em}} = q_{\mathrm{em}}^{-1}$, instead of $q_{\mathrm{em}}^{-2}$ as outlined above. This is the case for direct symmetry verification~\cite{mcardleErrorMitigatedDigitalQuantum2019,bonet-monroigLowcostErrorMitigation2018}.

We have not discussed how $\vec{p}_{\mathrm{rsp}}$ and $q_{\mathrm{em}}$ are obtained and the corresponding sampling overheads. This relies on the other key step in QEM, \emph{noise calibration}, which will be discussed in detail in \cref{sec:str_cali}. By combining the results from noise calibration and response measurement, we can understand how the calibrated noise components in the circuit affect the output observable and thus construct an estimator unaffected by these error components. A linear error-mitigated estimator is constructed from a linear combination of response measurement circuits with the weighting $\vec{p}_{\mathrm{rsp}}$ and/or the normalisation factor $q_{\mathrm{em}}$ given by noise calibration.

\section{Faults in the circuit}\label{sec:circ_fault}
From \cref{eqn:fid_boost,eqn:linear_samp_cost}, we see that the fidelities of the error-mitigated state and the unmitigated state are deeply related to the performance of the QEM method. In this section, we will study the distributions of faults in the circuit, which can then be used to estimate fidelities and analytically deduce the performance of the different QEM techniques under some practical assumptions. 

Let us assume there are $M$ \emph{locations} in the circuit where an error can occur. When a given location fails, it means the error associated with that location has occurred, and we call it a \emph{fault} in the circuit. Assuming the error probability at every location is $p$, the \emph{circuit fault rate}, which is the average number of faults occurring in each circuit run, is simply $\lambda = Mp$. Note that the circuit fault rate $\lambda$ can be greater than $1$. If $\lambda$ is too big, then there will be many faults in each circuit run such that the circuit would be too noisy to extract any useful information out of it. On the other hand, when the circuit fault rate is very low $\lambda \ll 1$, it would mean that our circuit runs almost perfectly, and no measures need to be taken to counter the noise in the circuit. For practical NISQ application, it is more realistic to assume that we have a finite circuit fault rate $\lambda \sim 1$ and the size of the circuit needed is beyond dozens of gates: $M \gg 1$. Under such an assumption, for \emph{Markovian} noise, the probability that exactly $\ell$ faults occur in a given circuit run can be approximated from a binomial distribution to a Poisson distribution:
\begin{align}\label{eqn:fault_distribution}
    \prob{\abs{\mathbb{L}} = \ell} = e^{-\lambda}\frac{\lambda^\ell }{\ell!}.
\end{align}
Here $\mathbb{L}$ is the set of faults that occur in a given circuit run, which is called a \emph{fault path}. Using Le Cam's theorem, the Poisson fault distribution here will also apply to the case in which different locations have different error rates, with $\lambda$ being the circuit fault rate that is the sum of the error rates of all locations. Note that fault locations need not be local in space, thus it is possible for our model to account for spatially correlated errors.

The noisy output state we have can be written as:
\begin{align}\label{eqn:noisy_state_distri}
    \rho_{\lambda} &= \sum_{\ell = 0}^{\infty} \prob{\abs{\mathbb{L}} = \ell} \rho_{\abs{\mathbb{L}} = \ell} =  e^{-\lambda} \sum_{\ell = 0}^{\infty} \frac{\lambda^\ell }{\ell!} \rho_{\abs{\mathbb{L}} = \ell}
\end{align}
where $\rho_{\abs{\mathbb{L}} = \ell}$ is the average output state when there are $\ell$ faults in the circuit. Note that $ \rho_{\abs{\mathbb{L}} = 0} = \rho_0$.

We can think of $\rho_{\abs{\mathbb{L}} = \ell}$ as the erroneous state whose error is the uniform incoherent mixture of all $\ell$th order faults. If we make the simple assumption that such an incoherent mixture of faults will send the state to an orthogonal erroneous state, we have
\begin{align}\label{eqn:ortho_error}
    \rho_0\rho_{\abs{\mathbb{L}} = \ell} = \rho_{\abs{\mathbb{L}} = \ell}\rho_0 = 0 \quad \forall\ell \neq 0.
\end{align}
In this way, using \cref{eqn:noisy_state_distri}, the fidelity of the noisy output state is simply the probability that no faults have occurred:
\begin{align}\label{eqn:fid_approx}
    \Tr(\rho_0\rho_{\lambda}) =  \prob{\abs{\mathbb{L}} = 0} = e^{-\lambda}.
\end{align}
In practice for a given circuit, the fraction of fault paths that are benign is usually small, and thus this is usually a good order-of-magnitude estimate for the fidelity even if our assumption of `orthogonal' errors does not hold. Hence, we see that the fidelity of the circuit decays exponentially against the circuit fault rate. This is also the reason why QEM is mainly limited by the overall circuit fault rate instead of the local gate error rate, unlike QEC.

\section{Gate Error Mitigation}\label{sec:gate_error_mitigation}
Here we will talk about QEM schemes that make use of information about the gate errors (faults) in the circuit. This can be either the full noise characterisation in the case of probabilistic error cancellation or simply the overall circuit fault rate (determined by the gate error rate) in the case of zero-noise extrapolation. 

\subsection{Probabilistic Error Cancellation}
To illustrate the concepts of probabilistic error cancellation~\cite{temmeErrorMitigationShortDepth2017}, let us first look at a simple example. If the ideal state $\rho_0$ undergo dephasing noise with probability $p$ and turn into the noisy state $\rho_{p}$, we have $\rho_{p} = (1-p)\rho_0 + p Z \rho_0 Z $. We can rewrite this as:
\begin{align*}
    \rho_0 = (1+ \alpha) \rho_p - \alpha Z \rho_p Z
\end{align*}
with $\alpha = \frac{p}{1-2p}$. In this equation, we have effectively `inverted' the dephasing channel by taking the linear sum of the original noisy state $\rho_p$ and the noisy state with an additional $Z$ gate $Z \rho_p Z$. We can perform a similar `inversion' to a different error channel by taking a linear combination of states with different gates added. This will work even if the gates we add are noisy, as long as these gates can form a basis for representing the error channels~\cite{temmeErrorMitigationShortDepth2017,endoPracticalQuantumError2018}. 

Generalising the above arguments, we can `invert' the error channels at different fault locations in the primary circuit by taking the linear sum of a set of circuits with different gate additions at different fault locations. The output states of this set of circuits are denoted as $\vec{\rho}_{\mathrm{rsp}}$. By taking their linear sum and `inverting' all the error channels, we will obtain the ideal state as our error-mitigated state:
\begin{align}\label{eqn:quasi_em_state}
    \rho_{\mathrm{em}} = \rho_0 = \vec{\alpha} \cdot \vec{\rho}_{\mathrm{rsp}}.
\end{align}
Note that the coefficients $\vec{\alpha}$ are calculated using our knowledge of the forms of the error channels in the circuit~\cite{temmeErrorMitigationShortDepth2017,endoPracticalQuantumError2018}.

Using \cref{eqn:fid_boost,eqn:fid_approx}, the fidelity boost of probabilistic error cancellation is simply
\begin{align}\label{eqn:quasi_fid_boost}
    B_{\mathrm{em}} = p_{\mathrm{em}}^{-1} = \Tr(\rho_0\rho_{\lambda})^{-1} = e^{\lambda},
\end{align}

To obtain the sampling overhead, we can rewrite \cref{eqn:quasi_em_state} as
\begin{align}\label{eqn:quasi_state_extract}
    \frac{1}{A}\rho_0 = \sum_i\frac{\abs{\alpha_i}}{A}\sign(\alpha_i)\rho_{\mathrm{rsp},i}
\end{align}
with $A = \sum_i\abs{\alpha_i}$. 

Comparing \cref{eqn:quasi_state_extract} to \cref{eqn:em_extraction}, we see that in the case of probabilistic error cancellation, we can construct a set of response measurement circuits that can directly measure $\sign(\alpha_i)\Tr(O\rho_{\mathrm{rsp},i})$ for all $i$, which can be viewed as circuits with the effective output `states' $\sigma_{\mathrm{rsp}, i} = \sign(\alpha_i)\rho_{\mathrm{rsp},i}$. We will perform these circuits according to the probability distribution $p_{\mathrm{rsp}, i}  = \frac{\abs{\alpha_i}}{A}$, which will yield a resultant output `state' $q_{\mathrm{em}} \rho_{\mathrm{em}}$ that is a fraction of the error-mitigated state (note that $\rho_{\mathrm{em}} = \rho_0$) with
\begin{align*}
    q_{\mathrm{em}} = \frac{1}{A}.
\end{align*}
Using the fault distribution in \cref{sec:circ_fault} and using arguments in Ref.~\cite{endoPracticalQuantumError2018,caiMultiexponentialErrorExtrapolation2021}, we can obtain $ A = \sum_i\abs{\alpha_i} \sim e^{2\lambda}$. Hence, using \cref{eqn:linear_samp_cost}, the sampling overhead for probabilistic error cancellation is
\begin{align}\label{eqn:quasi_samp_cost}
    C_{\mathrm{em}} \sim q_{\mathrm{em}}^{-2} = A^{-2} \sim e^{4\lambda}.
\end{align}

Using \cref{eqn:extraction_rate}, the extraction rate for probabilistic error cancellation is:
\begin{align*}
    r_{\mathrm{em}} = \frac{q_{\mathrm{em}}}{p_{\mathrm{em}}} = e^{-\lambda}.
\end{align*}

Instead of removing all the noise, we can instead try to suppress the noise only partially to the circuit fault rate $\lambda_{\mathrm{em}}$~ \cite{caiMultiexponentialErrorExtrapolation2021}. Assuming we are reducing the noise levels to various degree at different fault locations without changing the form of the errors, we have $\Tr(\rho_{0}\rho_{\mathrm{em}}) = e^{-\lambda_{\mathrm{em}}}$ and using arguments in Ref.~\cite{caiMultiexponentialErrorExtrapolation2021}, we see that we have the same expressions as before but with the circuit fault rate $\lambda$ replaced by the \emph{reduction} in the circuit fault rate $\lambda - \lambda_{\mathrm{em}}$:
\begin{align*}
    B_{\mathrm{em}} &= e^{\lambda - \lambda_{\mathrm{em}}}\\
    C_{\mathrm{em}} &\sim e^{4\left(\lambda - \lambda_{\mathrm{em}}\right)}\\
    r_{\mathrm{em}} &=  e^{-\left(\lambda - \lambda_{\mathrm{em}}\right)}.
\end{align*}

\subsection{Zero-noise Extrapolation}
In zero-noise extrapolation, we will view the noisy expectation value $\Tr(O\rho_{\lambda})$ as a function of the circuit fault rate $\lambda$. Suppose the smallest circuit fault rate we can achieve is $\lambda$, and we can obtain the noisy expectation values $\Tr(O\rho_{\lambda_i})$ at different \emph{boosted} error rates $\lambda_i$ with $\lambda_i \geq \lambda$, we can then try to estimate the zero-noise expectation value $\Tr(O\rho_{0})$ by fitting a function to these data points. We will assume we have probed at $n$ different circuit fault rates $\{\lambda_1, \cdots, \lambda_n\}$, with $\lambda_i > \lambda_{i-1}$ and $\lambda_1 = \lambda$. If the maximum circuit fault rate $\lambda_n$ is small, then we can apply Richardson extrapolation as discussed in Ref.~\cite{temmeErrorMitigationShortDepth2017} with the error-mitigated expectation value given by:
\begin{align}
    \Tr(O\rho_{\mathrm{em}}) = \sum_{i = 1}^{n} \gamma_i\Tr(O\rho_{\lambda_i})  = \Tr(O\rho_{0}) + \order{\lambda^n}
\end{align}
where
\begin{align*}
    \gamma_i = \prod_{\substack{k = 1\\ k \neq i}}^n \frac{\lambda_k}{\lambda_k - \lambda_i}, \quad \sum_{i = 1}^n \gamma_i = 1.
\end{align*}
We can see that leading order bias is suppressed to $\order{\lambda^n}$, but it is hard to obtain an analytical expression for the fidelity boost. 

To illustrate the performance of extrapolation-type techniques, in this section we will study another extrapolation strategy called \emph{analytical extrapolation}, for which we can analytically derive its fidelity boost and sampling overhead. It can be viewed as performing Richardson extrapolation for the function $ \Tr(O\rho_{\lambda}) e^{\lambda}$ instead of $\Tr(O\rho_{\lambda})$ and its error-mitigated expectation value is given by:
\begin{align}\label{eqn:zne_em_est}
    \Tr(O\rho_{\mathrm{em}}) &= \frac{1}{A}\sum_{i = 1}^{n} \alpha_i \Tr(O\rho_{\lambda_i}) 
\end{align}
where 
\begin{align*}
    \alpha_i &=  \gamma_i e^{\lambda_i} ,\quad A = \sum_{i = 1}^{n} \alpha_i = \sum_{i = 1}^{n} \gamma_i e^{\lambda_i} .
\end{align*}
Here $A$ is the normalising factor to ensure we have $\Tr(\rho_{\mathrm{em}}) = 1$ with $O = I$. Note that $A$ is simply the Richardson zero-noise estimator for the function $e^{\lambda}$, hence it is simply here to remove the approximation error caused by the extra $e^{\lambda}$ factor in the function we fit. 

\cref{eqn:zne_em_est} can be viewed as obtaining the expectation value for the error-mitigated state 
\begin{align}\label{eqn:zne_em_state}
    \rho_{\mathrm{em}} &= \frac{1}{A}\sum_{i = 1}^{n} \alpha_i \rho_{\lambda_i}.
\end{align}
Using \cref{eqn:noisy_state_distri} and arguments in \cref{sec:ana_extra_state}, we see that the state can be written as:
\begin{align}\label{eqn:zne_em_state_simp}
    \rho_{\mathrm{em}} &= \frac{1}{A}  \left(\rho_{0} +  \sum_{\ell = n}^{\infty}  c_{\ell}(\vec{\lambda}) \frac{\lambda^\ell}{\ell!}\rho_{\abs{\mathbb{L}} = \ell}\right)
\end{align}
with $c_{\ell}(\vec{\lambda}) = \sum_{i = 1}^{n} \gamma_i \left(\nicefrac{\lambda_i}{\lambda}\right)^\ell$. i.e. our extrapolation removes all the fault paths with less than $n$ faults. 

Using the assumption in \cref{eqn:ortho_error}, we can obtain the fidelity of the error-mitigated state to be:
\begin{align}\label{eqn:zne_em_fid}
    \Tr(\rho_0\rho_{\mathrm{em}}) = A^{-1}.
\end{align}
Now using \cref{eqn:fid_boost,eqn:fid_approx}, we can obtain the fidelity boost using analytical extrapolation to be
\begin{align}\label{eqn:zne_fid_boost}
    B_{\mathrm{em}} = p_{\mathrm{em}}^{-1} = \frac{\Tr(\rho_0\rho_{\mathrm{em}})}{\Tr(\rho_0\rho_{\lambda})} = \frac{e^{\lambda}}{A} = \frac{e^{\lambda}}{\sum_i\gamma_i e^{\lambda_i} }.
\end{align}

Note that in order to have a fidelity boost $B_{\mathrm{em}}$ larger than $1$, we need to have $A \leq e^{\lambda}$, which implies none of the $\abs{\gamma_i}$ and $\lambda_i$ can be too big. This in turn means that we cannot have the gap between probed error rates $\abs{\lambda_i - \lambda_j}$ tends to zero or infinity for any $i$ and $j$, giving us restrictions on the set of error rates we can probe.

To obtain the sampling overhead of analytical extrapolation, we can rewrite \cref{eqn:zne_em_state} as
\begin{align}\label{eqn:zne_state_extract}
    \frac{A}{A_{\mathrm{abs}}} \rho_{\mathrm{em}}&= \sum_{i = 1}^{n} \frac{\abs{\alpha_i}}{A_{\mathrm{abs}}} \sign(\alpha_i) \rho_{\lambda_i} 
\end{align}
with $A_{\mathrm{abs}} = \sum_i\abs{\alpha_i} = \sum_i\abs{\gamma_i} e^{\lambda_i}$.

Comparing \cref{eqn:zne_state_extract} to \cref{eqn:em_extraction}, we see that in the case of analytical extrapolation, we can construct a set of response measurement circuits that can directly measure $\sign(\alpha_i)\Tr(O\rho_{\lambda_i} )$ for all $i$, which can be viewed as circuits with the effective output `states' $\sigma_{\mathrm{rsp}, i} = \sign(\alpha_i)\rho_{\lambda_i}$. We will perform these circuits according to the probability distribution $p_{\mathrm{rsp}, i}  = \abs{\alpha_i}/A_{\mathrm{abs}}$, which will yield a resultant output `state' $q_{\mathrm{em}} \rho_{\mathrm{em}}$ that is a fraction of the error-mitigated state with
\begin{align*}
    q_{\mathrm{em}} = \frac{A}{A_{\mathrm{abs}}}.
\end{align*} 
Hence, using \cref{eqn:linear_samp_cost}, the sampling cost for analytical extrapolation is
\begin{align}\label{eqn:zne_samp_cost}
    C_{\mathrm{em}} \sim q_{\mathrm{em}}^{-2} = \frac{A_{\mathrm{abs}}^2}{A^2} = \left(\frac{\sum_i\abs{\gamma_i} e^{\lambda_i} }{\sum_i\gamma_i e^{\lambda_i} }\right)^2.
\end{align}

Using \cref{eqn:extraction_rate}, the extraction rate for analytical extrapolation is:
\begin{align}\label{eqn:zne_extraction_rate}
    r_{\mathrm{em}} = \frac{q_{\mathrm{em}}}{p_{\mathrm{em}}} = \frac{e^{\lambda}}{A_{\mathrm{abs}}}.
\end{align}

From \cref{sec:ana_extra_coef}, we see that for the simple case of $\lambda_n = n\lambda$, we have:
\begin{align*}
    B_{\mathrm{em}} &= \frac{e^{\lambda}}{A} = \frac{e^{\lambda}}{\left(e^{\lambda} - 1\right)^n + 1}\\
    C_{\mathrm{em}} &\sim \left(\frac{A_{\mathrm{abs}}}{A}\right)^2 = \left(\frac{\left(e^{\lambda}+1\right)^n - 1}{\left(e^{\lambda} - 1\right)^n + 1}\right)^2\\
    r_{\mathrm{em}} &= \frac{e^{\lambda}}{A_{\mathrm{abs}}} = \frac{e^{\lambda}}{\left(e^{\lambda}+1\right)^n - 1}.
\end{align*}

\section{State Error Mitigation}\label{sec:state_error_mitigation}
Instead of mitigating the damages of noise in a generic gate-based manner, ideally we would want to mitigate the damages of noise inflicted on our particular state. It is hard to characterise such damages since we do not have the full description of our state. Thus instead we turn to the known constraints of the states, which are usually given by the underlying physical problem we try to solve. Here we will look at three types of constraints placed on our state: constraints on the target observable of the state, symmetry constraints and purity constraints.  

\subsection{Subspace Expansion}
In subspace expansion, we are given a set of expansion basis $\vec{G}$ whose linear combination according to the probability distribution $\vec{w}$ will give us the expansion operator:
\begin{align}\label{eqn:expand_op}
    \Gamma_{\vec{w}} = \vec{w}\cdot \vec{G}.
\end{align}
In subspace expansion, we will identify a probability distribution $\vec{w}$ such that the subspace-expanded state:
\begin{align}\label{eqn:expand_em_state}
    \rho_{\mathrm{em}} = \frac{\Gamma_{\vec{w}} \rho_{\lambda} \Gamma_{\vec{w}}^\dagger}{\Tr(\Gamma_{\vec{w}}^\dagger\Gamma_{\vec{w}}\rho_{\lambda})}
\end{align}
will be the error-mitigated state that can produce results with less estimation bias. By changing $\vec{w}$, the state $\rho_{\mathrm{em}}$ is moving within a subspace and thus the name subspace expansion. In the original proposal~\cite{mccleanHybridQuantumclassicalHierarchy2017}, we are trying to find the ground state energy. Thus the optimal $\vec{w}$ is found by minimising the energy within the subspace. This in a sense can be viewed as a \emph{constraint} that our resultant state must have the minimal energy as given by the problem we try to solve. Of course, if we are targetting other observables, then our constraint and optimisation strategy will also change accordingly. 

The fidelity boost achieved can be obtained using \cref{eqn:fid_boost}:
\begin{align}\label{eqn:expand_fid_boost}
    B_{\mathrm{em}} = p_{\mathrm{em}}^{-1}= \frac{\Tr(\rho_0\rho_{\mathrm{em}})}{\Tr(\rho_0\rho_{\lambda})} = \frac{\Tr(\Gamma_{\vec{w}}^\dagger\rho_0\Gamma_{\vec{w}} \rho_{\lambda}) }{\Tr(\rho_0\rho_{\lambda})\Tr(\Gamma_{\vec{w}}^\dagger\Gamma_{\vec{w}}\rho_{\lambda})}
\end{align}

Using \cref{eqn:expand_op}, we can rewrite \cref{eqn:expand_em_state} as:
\begin{align}\label{eqn:expand_state_extract}
    \Tr(\Gamma_{\vec{w}}^\dagger\Gamma_{\vec{w}}\rho_{\lambda}) \rho_{\mathrm{em}} = \sum_{j,k}w_jw_k G_j \rho_{\lambda} G_k^\dagger.
\end{align}

We can construct a set of response measurement circuits that can directly measure $\Tr(G_k^\dagger OG_j \rho_{\lambda}) = \Tr(O G_j \rho_{\lambda} G_k^\dagger) $ for all $j,k$, which essentially output the effective `states' $\sigma_{\mathrm{rsp}, (j, k)} = G_j \rho_{\lambda} G_k^\dagger$. The R.H.S. of \cref{eqn:expand_state_extract} is simply an ensemble of such states following the probability distribution $p_{\mathrm{rsp}, (j, k)}  = w_jw_k$. For the L.H.S. of \cref{eqn:expand_state_extract}, by comparing to \cref{eqn:em_extraction}, we see that this is simply a fraction of the error-mitigated state $q_{\mathrm{em}} \rho_{\mathrm{em}}$ with
\begin{align}\label{eqn:expand_em_fraction}
    q_{\mathrm{em}} = \Tr(\Gamma_{\vec{w}}^\dagger\Gamma_{\vec{w}}\rho_{\lambda}).
\end{align}
Hence, using \cref{eqn:linear_samp_cost}, the sampling cost for subspace expansion is
\begin{align}\label{eqn:expand_samp_cost}
    C_{\mathrm{em}} \sim q_{\mathrm{em}}^{-2} = \Tr(\Gamma_{\vec{w}}^\dagger\Gamma_{\vec{w}}\rho_{\lambda})^{-2}.
\end{align}

Using \cref{eqn:extraction_rate}, the extraction rate for subspace expansion is:
\begin{align}\label{eqn:expand_extract_rate}
    r_{\mathrm{em}} = \frac{q_{\mathrm{em}}}{p_{\mathrm{em}}} = \frac{\Tr(\Gamma_{\vec{w}}^\dagger\rho_0\Gamma_{\vec{w}} \rho_{\lambda}) }{\Tr(\rho_0\rho_{\lambda})}.
\end{align}

\subsection{Symmetry Verification}
In symmetry verification, we know that the ideal output state $\rho_0$ is `stabilised' by a \emph{group} of symmetry operators $\mathbb{S}$:
\begin{align}\label{eqn:sym_cond}
    S \rho_0 = \rho_0 S = \rho_0\quad \forall S \in \mathbb{S}
\end{align}
Hence, the ideal state lives within the symmetry subspace defined by the projection operator:
\begin{align}\label{eqn:proj_op}
    \Pi_{\mathbb{S}} = \frac{1}{\abs{\mathbb{S}}}\sum_{S \in \mathbb{S}} S.
\end{align}
We can try to remove the noise in the noisy state $\rho_{\lambda}$ that violates the symmetry condition in \cref{eqn:sym_cond} by projecting the noisy state back into the symmetry subspace
\begin{align}\label{eqn:sym_em_state}
    \rho_{\mathrm{em}} = \frac{\Pi_{\mathbb{S}} \rho_{\lambda} \Pi_{\mathbb{S}}}{\Tr(\Pi_{\mathbb{S}} \rho_{\lambda})}.
\end{align}
Here we have used the fact that the projection operator $\Pi_{\mathbb{S}}$ is Hermitian and idempotent. Compared to \cref{eqn:expand_em_state}, we see that this is simply subspace expansion with the set of expansion basis $\vec{G}$ being the set of symmetry operators $\mathbb{S}$, and the expansion probability distribution $\vec{w}$ is \emph{uniform}.

Using \cref{eqn:expand_fid_boost,eqn:expand_samp_cost,eqn:expand_extract_rate} with $\Gamma_{\vec{w}} = \Gamma_{\vec{w}}^\dagger =\Pi_{\mathbb{S}}$ and $\Pi_{\mathbb{S}} \rho_0 = \rho_0\Pi_{\mathbb{S}} = \rho_0$, we can obtain the fidelity boost, sampling cost and extraction rate of symmetry verification to be:
\begin{align}
    B_{\mathrm{em}} &= p_{\mathrm{em}}^{-1} =  \Tr(\Pi_{\mathbb{S}} \rho_{\lambda})^{-1} \label{eqn:sym_fid_boost}\\
    C_{\mathrm{em}} &\sim q_{\mathrm{em}}^{-2} = \Tr(\Pi_{\mathbb{S}} \rho_{\lambda})^{-2} \label{eqn:sym_samp_cost}\\
    r_{\mathrm{em}} &= \frac{q_{\mathrm{em}}}{p_{\mathrm{em}}} = 1\label{eqn:sym_extract_rate}.
\end{align}
From the extraction rate, we see that $q_{\mathrm{em}} = p_{\mathrm{em}}$, i.e. all error-mitigated components (components with the right symmetry) can be effectively extracted via symmetry verification. 

We can gain more insights by considering the case of Pauli symmetry and we have a circuit consisting of symmetry-preserving components with faults only occurring in between them. For any given symmetry, we will assume any fault in the circuit will either commute or anti-commute with it, which means that the fault will be undetectable (commute) or detectable (anti-commute) for that particular symmetry. Denoting the fraction of faults detectable by the symmetry $S$ as $f_S$ (a given fault can be detectable by multiple symmetries), then following the fault distribution in \cref{sec:circ_fault} and arguments in Ref.~\cite{caiQuantumErrorMitigation2021a}, we can obtain:
\begin{align*}
    \Tr(\Pi_{\mathbb{S}} \rho_{\lambda}) = \frac{1}{\abs{\mathbb{S}}}\sum_{S \in \mathbb{S}} e^{-2f_S\lambda} 
\end{align*}
i.e. it is the function $e^{-2f_S\lambda}$ averaged over all symmetry operators. Note that $I$ is always one of the symmetry operators and it can detect zero fraction of faults $f_{I} = 0$.

Using Jensen's inequality, we have:
\begin{align*}
    \Tr(\Pi_{\mathbb{S}} \rho_{\lambda}) \geq e^{-2f_{\mathbb{S}}\lambda} 
\end{align*}
where
\begin{align*}
    f_{\mathbb{S}} = \frac{1}{\abs{\mathbb{S}}}\sum_{S \in \mathbb{S}} f_S 
\end{align*}
is the average fraction of faults that are detectable by our group of symmetry.

Using \cref{eqn:sym_fid_boost}, the fidelity boost for symmetry verification is:
\begin{align*}
    B_{\mathrm{em}} = p_{\mathrm{em}}^{-1} = \abs{\mathbb{S}} \left(\sum_{S \in \mathbb{S}} e^{-2f_S\lambda}\right)^{-1} \leq e^{2f_{\mathbb{S}}\lambda} 
\end{align*}
and the sampling overhead using \cref{eqn:sym_samp_cost,eqn:sym_extract_rate} is simply:
\begin{align*}
    C_{\mathrm{em}}  = q_{\mathrm{em}}^{-2} = p_{\mathrm{em}}^{-2} = \abs{\mathbb{S}}^2 \left(\sum_{S \in \mathbb{S}} e^{-2f_S\lambda}\right)^{-2} \leq e^{4f_{\mathbb{S}}\lambda}.
\end{align*}

\subsection{Purification-based Methods}
In purification-based QEM~\cite{hugginsVirtualDistillationQuantum2021,koczorExponentialErrorSuppression2021}, by knowing our ideal state $\rho_0$ is a pure state and it is the \emph{dominant eigenvector} within the noisy state $\rho_{\lambda}$, we can amplify the ideal state component by constructing the purified state 
\begin{align}\label{eqn:pur_em_state}
    \rho_{\mathrm{em}} = \frac{\rho_{\lambda}^n}{\Tr(\rho_{\lambda}^n)}.
\end{align}
If the dominant eigenvector is not exactly the ideal state, it will lead to a coherent mismatch which will bound the performance achieved using purification-based methods, but this is usually not the bottleneck in practical applications~\cite{koczorDominantEigenvectorNoisy2021}. Assuming that the ideal state is the dominant eigenvector, we can decompose the noisy state into
\begin{align}\label{eqn:pure_decomp}
    \rho_{\lambda} = \Tr(\rho_0\rho_{\lambda}) \rho_0 + (1- \Tr(\rho_0\rho_{\lambda})) \rho_{\epsilon}
\end{align}
with $\rho_{\epsilon}$ being the noisy component. Note that we have $\rho_{\epsilon}\rho_{0} = \rho_{0}\rho_{\epsilon} = 0$.

Using \cref{eqn:pur_em_state,eqn:pure_decomp}, we can write the error-mitigated state in purification-based QEM as:
\begin{align*}
    \rho_{\mathrm{em}} = \frac{\rho_{\lambda}^n}{\Tr(\rho_{\lambda}^n)} = \frac{ \rho_0 +  \left(\Tr(\rho_0\rho_{\lambda})^{-1}-1\right)^n \rho_\epsilon^n}{ 1 + \left(\Tr(\rho_0\rho_{\lambda})^{-1}-1\right)^n\Tr(\rho_\epsilon^n)}.
\end{align*}
Then using \cref{eqn:fid_boost}, we can obtain the fidelity boost of purification-based QEM:
\begin{align}\label{eqn:pur_fid_boost}
    B_{\mathrm{em}} = p_{\mathrm{em}}^{-1} = \frac{\Tr(\rho_0\rho_{\mathrm{em}})}{\Tr(\rho_0\rho_{\lambda})}  &= \frac{\Tr(\rho_0\rho_{\lambda})^{-1}}{ 1 + \left(\Tr(\rho_0\rho_{\lambda})^{-1}-1\right)^n\Tr(\rho_\epsilon^n)}\nonumber\\
    &\geq \frac{\Tr(\rho_0\rho_{\lambda})^{-1}}{ 1 + \left(\Tr(\rho_0\rho_{\lambda})^{-1}-1\right)^n} 
\end{align}
where we have used $\Tr(\rho_\epsilon^n) \leq 1$.

To obtain the sampling overhead of purification-based QEM, we can rewrite \cref{eqn:pur_em_state} as 
\begin{align}\label{eqn:pur_state_extract}
    \Tr(\rho_{\lambda}^n) \rho_{\mathrm{em}} = \rho_{\lambda}^n.
\end{align}
In Ref.~\cite{koczorExponentialErrorSuppression2021,hugginsVirtualDistillationQuantum2021}, we see that we can measure $\Tr(O\rho_{\lambda}^n)$ using $n$ copies of the noisy machines by measuring the derangement operator $D$ among the copies and the observable $O$ on the first copy:
\begin{align}\label{eqn:multi-copy_pur}
    \Tr(O\rho_{\lambda}^n) = \Tr(O_1D\rho_{\lambda}^{\otimes n}).
\end{align}
This will be called \emph{multi-copy purification} and it can be viewed as a response measurement circuit with the effective output `states' $\rho_{\lambda}^n$. In fact, this is the only response measurement circuit we need to extract a fraction of the error-mitigated state $q_{\mathrm{em}} \rho_{\mathrm{em}}$  with
\begin{align*}
    q_{\mathrm{em}} =  \Tr(\rho_{\lambda}^n) = \Tr(\rho_0\rho_{\lambda})^n \left[1 + \left(\Tr(\rho_0\rho_{\lambda})^{-1} - 1\right)^n\Tr(\rho_\epsilon^n)\right].
\end{align*} 
Hence, using \cref{eqn:linear_samp_cost}, the sampling cost for multi-copy purification is
\begin{align}\label{eqn:pur_samp_cost}
    C_{\mathrm{em}} \sim q_{\mathrm{em}}^{-2} &= \frac{\Tr(\rho_0\rho_{\lambda})^{-2n}}{\left[1 + \left(\Tr(\rho_0\rho_{\lambda})^{-1} - 1\right)^n\Tr(\rho_\epsilon^n)\right]^2}\nonumber\\
    &\geq \frac{\Tr(\rho_0\rho_{\lambda})^{-2n}}{\left[1 + \left(\Tr(\rho_0\rho_{\lambda})^{-1} - 1\right)^n\right]^2}
\end{align}
where we have also used $\Tr(\rho_\epsilon^n) \leq 1$.

Using \cref{eqn:extraction_rate}, the extraction rate for multi-copy purification is:
\begin{align*}
    r_{\mathrm{em}} = \frac{q_{\mathrm{em}}}{p_{\mathrm{em}}} = \Tr(\rho_0\rho_{\lambda})^{n-1}.
\end{align*}

Our model of fault distribution in \cref{sec:circ_fault} will not lead to any coherent mismatch. Using this model, we have $\Tr(\rho_0\rho_{\lambda}) = e^{-\lambda}$, which gives:
\begin{align*}
     B_{\mathrm{em}} &= p_{\mathrm{em}}^{-1} = \frac{e^{\lambda}}{ 1 + \left(e^{\lambda}-1\right)^n \Tr(\rho_\epsilon^n)}\\
     C_{\mathrm{em}}  &\sim q_{\mathrm{em}}^{-2} =  \frac{e^{2n\lambda}}{\left(1+ \left(e^{\lambda}-1\right)^n\Tr(\rho_\epsilon^n)\right)^2}\\
     r_{\mathrm{em}} &= \frac{q_{\mathrm{em}}}{p_{\mathrm{em}}} = e^{-(n-1)\lambda}.
\end{align*}

\section{Discussion}\label{sec:comp_comb}
\subsection{Comparison}\label{sec:comp}
\begin{table*}[tbhp!]
    \centering
    \begingroup
    \renewcommand{\arraystretch}{1.5}
    \setlength{\tabcolsep}{1.3em}
    \begin{tabularx}{\textwidth}{p{5em} p{9em} p{6.5em} p{-2em} p{9.5em} p{8.5em}}\toprule
        &\multicolumn{2}{c}{Gate Error Mitigation}&&\multicolumn{2}{c}{State Error Mitigation}\\\cline{2-3}\cline{5-6}
        & Probabilistic Error Cancellation  & Analytical Extrapolation && Pauli Symmetry Verification& Multi-copy Purification\\\hline
        Key Quantities & Circuit fault rate after mitigation: $\lambda_{\mathrm{em}}$ & Number of data points: $n$ & & Set of symmetries: $\mathbb{S}$; Fraction of errors detectable by $S \in \mathbb{S}$: $f_S$ & Number of device copies: $n$\\
        Fidelity boost ($B_{\mathrm{em}}$) &\large$e^{\lambda - \lambda_{\mathrm{em}}}$ &\large$\frac{e^{\lambda}}{\left(e^{\lambda} - 1\right)^n + 1}$& &\large$\frac{\abs{\mathbb{S}}}{\sum_{S \in \mathbb{S}} e^{-2f_S\lambda}}$ &\large$\frac{e^{\lambda}}{ 1 + \left(e^{\lambda}-1\right)^n \Tr(\rho_\epsilon^n)}$\\
        Sampling overhead ($C_{\mathrm{em}}$) &\large$e^{4\left(\lambda - \lambda_{\mathrm{em}}\right)}$ &\large$\left(\frac{\left(e^{\lambda}+1\right)^n - 1}{\left(e^{\lambda} - 1\right)^n + 1}\right)^2$& &\large$\left(\frac{\abs{\mathbb{S}}}{\sum_{S \in \mathbb{S}} e^{-2f_S\lambda}}\right)^{2}$ &\large$\left(\frac{e^{n\lambda}}{ 1 + \left(e^{\lambda}-1\right)^n \Tr(\rho_\epsilon^n)}\right)^2$\\
        Extraction rate ($r_{\mathrm{em}} = \nicefrac{B_{\mathrm{em}}}{\sqrt{C_{\mathrm{em}}}}$) &\large$e^{-\left(\lambda - \lambda_{\mathrm{em}}\right)}$  &\large$\frac{e^{\lambda}}{\left(e^{\lambda}+1\right)^n - 1}$& &\large$1$ &\large$e^{-(n-1)\lambda}$\\
        \botrule
    \end{tabularx}
    \endgroup
    \caption{Summary of fidelity boost, sampling overhead and extraction rate for different QEM methods. The unmitigated circuit fault rate is $\lambda$. }
    \label{tab:qem_schemes}
\end{table*}

We have summarised the fidelity boost, sampling overhead and extraction rate for different QEM methods in \cref{tab:qem_schemes}. 

First we realise that symmetry verification has the highest extraction rate achievable at $1$, i.e. it can extract all components of the error-mitigated (symmetry-verified) state out of the noisy state, so it is in a sense the most cost-effective. However, for a given problem and qubit encoding, the set of symmetry is fixed, and we can only achieve a fixed level of fidelity boost. On the other hand, by changing the remaining circuit fault rate $\lambda_{\mathrm{em}}$ in probabilistic error cancellation, the number of data points $n$ in analytical extrapolation and the number of copies $n$ in multi-copy purification, we can achieve a fidelity boost up to full error removal as shown in \cref{tab:qem_schemes}. Thus symmetry verification, though very cost-effective, is often not enough to be the \emph{only} QEM scheme to be employed in practical NISQ applications since the amount of noise it can remove is limited. Of course, we can try to circumvent this by employing qubit encodings that have more inherent symmetries~\cite{bravyiTaperingQubitsSimulate2017,jiangMajoranaLoopStabilizer2019,derbyCompactFermionQubit2021a}, but this will come with a larger qubit overhead and the complexity of the gates will also likely to increase such that the overall errors might actually increase. 

Probabilistic error cancellation can achieve the second-highest extraction rate:
\begin{align}\label{eqn:extract_rate_comp}
    e^{-\left(\lambda - \lambda_{\mathrm{em}}\right)} \geq e^{-(n-1)\lambda}\geq \frac{e^{\lambda}}{\left(e^{\lambda}+1\right)^n - 1} \quad \forall n \geq 1.
\end{align}
Thus, it should be our preferred method to tackle the noise that symmetry verification cannot remove. The optimality of probabilistic error cancellation is also found in Ref.~\cite{takagiFundamentalLimitationsQuantum2021}. However, applying probabilistic error cancellation to any fault location would require a complete description of the error channel at that location, which can be costly to calibrate and is often not available in practice or not known to high enough precision. 

Looking at analytical extrapolation and multi-copy purification, we realise that using the same number of data points as the number of copies $n$, multi-copy purification can achieve a higher fidelity boost:
\begin{align}\label{eqn:fid_bst_comp}
    \frac{e^{\lambda}}{ 1 + \left(e^{\lambda}-1\right)^n \Tr(\rho_\epsilon^n)} \geq \frac{e^{\lambda}}{ 1 + \left(e^{\lambda}-1\right)^n}
\end{align}
where we have used $\Tr(\rho_\epsilon^n) \leq 1$. Using the right-hand side of \cref{eqn:fid_bst_comp}, we see that both methods can achieve a fidelity boost of $B_{\mathrm{em}} \geq 1$ as long as $e^{-\lambda} \geq 1/2 \Rightarrow \lambda \lesssim 0.7$. This is to ensure that the ideal state $\rho_0$ will be the \emph{dominant component} in the unmitigated state $\rho_{\lambda}$ with $e^{-\lambda} \geq 1/2$ in \cref{eqn:pure_decomp}. In practice, the noisy component $\rho_\epsilon$ in \cref{eqn:pure_decomp} is usually far from pure and thus $\Tr(\rho_\epsilon^n)$ is expected to be small, allowing multi-copy purification to perform beyond the lower bound in \cref{eqn:fid_bst_comp}, and is able to work beyond the circuit fault rate of $\lambda \sim 0.7$. As shown in \cref{eqn:extract_rate_comp}, multi-copy extrapolation can also achieve a higher extraction rate than analytical extrapolation given the same $n$, and thus is more cost-effective. 

However, this is not to say that purification-based methods will outperform all extrapolation methods. For example, we have not talked about extrapolations that have more data points than free parameters~\cite{endoPracticalQuantumError2018,caiMultiexponentialErrorExtrapolation2021} since they do not fit in the linear QEM framework. We also have not taken into account the noise due to the additional circuit structure required for measuring the derangement operator in purification-based methods. Furthermore, when we have $n$ copies of the machines, instead of running one round of multi-copy purification, we can run $n$ rounds of other QEM methods in parallel. Of course, one may try to trade the number of copies with circuit depth by using state verification purification~\cite{obrienErrorMitigationVerified2021,huoDualstatePurificationPractical2021,caiResourceefficientPurificationbasedQuantum2021}. However, using circuits that are twice the depth of the primary circuit means that the time required to run one round of them, we can run two rounds of most of the other QEM schemes. Hence, a more illustrative metric for the total overhead of implementing a given QEM scheme might be the product of sampling overhead, qubit overhead and circuit runtime overhead.

\subsection{Combination}\label{sec:comb}
Since our framework views the process of QEM as extracting the error-mitigated state out of the original noisy state, it is natural to think that we can apply such extractions in sequence using different QEM methods. In such a way, we are essentially concatenating different QEM methods. The fidelity boost, sampling overhead and extraction rate of such multi-stage combined QEM will simply be the product of the fidelity boosts, sampling overheads and extraction rates achieved at individual stages. Note that implementing one QEM method on top of another is not always possible as we will see later. 

Since state error mitigation is implemented by making additional measurements on the output state, it would still work even if we are applying them to the effective output state coming out of a QEM scheme. Hence, we can directly apply state error mitigation on top of other QEM schemes. The expectation value of applying multi-copy purification and symmetry verification on top of the error-mitigated state $\rho_{\mathrm{em}}$ is
\begin{align}\label{eqn:comb_expval}
    \frac{\Tr(O\left(\Pi_{\mathbb{S}}\rho_{\mathrm{em}}\right)^{n})}{\Tr(\left(\Pi_{\mathbb{S}}\rho_{\mathrm{em}}\right)^{n})}
\end{align}
where we have assumed that our observable $O$ commute with the set of symmetries $\mathbb{S}$, which is usually the case in practice.

Using \cref{eqn:em_extraction,eqn:multi-copy_pur,eqn:proj_op}, the numerator in \cref{eqn:comb_expval} can be rewritten as:
\begin{align*}
    &\quad \Tr(O_1 D \left(\Pi_{\mathbb{S}}\rho_{\mathrm{em}}\right)^{\otimes n}) \\
    &= \Tr(O_1 D \left(\sum_{j = 1}^{\abs{\mathbb{S}}} \frac{S_j}{\abs{\mathbb{S}}} \frac{\sum_{k=1}^K p_{\mathrm{rsp},k} \sigma_{\mathrm{rsp},k}}{q_{\mathrm{em}}}\right)^{\otimes n})
\end{align*}
where $K$ is the size of the set of response measurement circuits for the error-mitigated state $\rho_{\mathrm{em}}$. It can be rewritten as 
\begin{align*}
    \Tr(O_1 D \left(\Pi_{\mathbb{S}}\rho_{\mathrm{em}}\right)^{\otimes n})
    & = \frac{1}{q_{\mathrm{em}}^{n}} \sum_{\vec{j},\vec{k}} \frac{p_{\vec{k}}}{\abs{\mathbb{S}}^n} \Tr(O_1 S_{\vec{j}} D  \sigma_{\vec{k}})
\end{align*}
where
\begin{gather*}
    S_{\vec{j}} = \bigotimes_{m = 1}^{n} S_{j_m},\quad p_{\vec{k}} = \prod_{m = 1}^{n} p_{\mathrm{rsp},k_m}, \quad \sigma_{\vec{k}} = \bigotimes_{k = 1}^{n} \sigma_{\mathrm{rsp},k_m}\\
    \vec{j} \in \{1,2,\cdots \abs{\mathbb{S}}\}^{\otimes n},\quad \vec{k} \in \{1,2,\cdots K\}^{\otimes n}.
\end{gather*}
Hence, to apply multi-copy purification and symmetry verification on top of the error-mitigated state, we will run the response measurement circuits that produce the output `state' $\sigma_{\vec{k}}$ with the probability $p_{\vec{k}}$. Then we will measure the derangement operator $D$ and a randomly chosen symmetry operator $S_{\vec{j}}$, along with the observable of interests $O$ on the first copy. The denominator in \cref{eqn:comb_expval} is obtained in the same way discussed above but with observable $I$ instead of $O$. Note that applying only multi-copy purification corresponds to the case of $\mathbb{S} = \{I\}$ and applying only symmetry verification corresponds to the case of $n = 1$. 

On the other hand, if we want to apply gate error mitigation on top of other QEM schemes, we must know how the other QEM schemes have affected our knowledge about the gate errors, which cannot always be done and would require more elaborate analysis (an example is in Ref.~\cite{caiMultiexponentialErrorExtrapolation2021}).

\section{Noise Calibration}\label{sec:str_cali}
As discussed in \cref{sec:qem_estimator}, we have not discussed how the value of $\vec{p}_{\mathrm{rsp}}$ and $q_{\mathrm{em}}$ are obtained and their corresponding sampling overheads. 

In gate error mitigation, we see that $\vec{p}_{\mathrm{rsp}}$ and $q_{\mathrm{em}}$ are given by our knowledge about the gate noise, which is obtained through standard gate noise calibration techniques~\cite{eisertQuantumCertificationBenchmarking2020}. These calibrations of gate noise can usually be performed during the device calibration stage. Thus the corresponding sampling overhead need not be counted towards the implementation cost of any specific computation tasks. 

In state error mitigation, we see that $\vec{p}_{\mathrm{rsp}}$ is known beforehand in the case of symmetry verification (uniform) and purification-based method (has just one response measurement circuit). However in these methods, $q_{\mathrm{em}}$ is unknown and can be obtained by measuring $I$ in place of $O$ in response measurement in \cref{eqn:rsp_cal}. Thus only additional measurements (sometimes post-processing) are needed for estimating $q_{\mathrm{em}}$, \emph{no additional circuit runs} are needed. The fact that $q_{\mathrm{em}}$ is obtained through estimation instead of being exactly known will not affect our order-of-magnitude sampling overhead estimate in \cref{eqn:linear_samp_cost} as shown in \cref{sec:state_samp_cost}.

In the case of subspace expansion, $\vec{p}_{\mathrm{rsp}}$ is obtained by solving the generalised eigenvalue problem for a matrix equation, with the entries in the matrices obtained by evaluating the subspace-expanded expectation value of $O$ and $I$ for different $\vec{p}_{\mathrm{rsp}}$. The overall sampling overhead, including evaluating the matrix elements, will depend on the expansion basis we choose, the problem we try to solve, etc., and cannot be analytically derived for the general case. 

Another class of methods to find the optimal $\vec{p}_{\mathrm{rsp}}$ for a given QEM method is learning-based methods~\cite{strikisLearningbasedQuantumError2021,czarnikErrorMitigationClifford2020,montanaroErrorMitigationTraining2021,googleaiquantumandcollaboratorsObservationSeparatedDynamics2020,loweUnifiedApproachDatadriven2021,bultriniUnifyingBenchmarkingStateoftheart2021}. In these methods, we use a series of \emph{classically simulable} circuits (e.g. Clifford circuits or free-fermion circuits) that are close to our primary circuits as the \emph{reference circuits} whose ideal expectation values can be exactly calculated. We can identify the optimal $\vec{p}_{\mathrm{rsp}}$ that can best mitigate the noise for the reference circuits through optimisation, and assume that the same optimal $\vec{p}_{\mathrm{rsp}}$ will be applicable to our primary circuit. The optimal $\vec{p}_{\mathrm{rsp}}$ is obtained through some optimisation algorithm and within each optimisation step we need to obtain the noisy expectation values of the reference circuits on real devices. The overall sampling overhead that includes the optimisation process has not yet been analytically derived. Note that noise calibration using reference circuits enables us to probe the strength of only the noise components that are damaging to our observable of interests instead of all noise components. 

\section{Conclusion}\label{sec:conclusion}
\begin{figure}[h!]
    \centering
    \includegraphics[width = 0.47\textwidth]{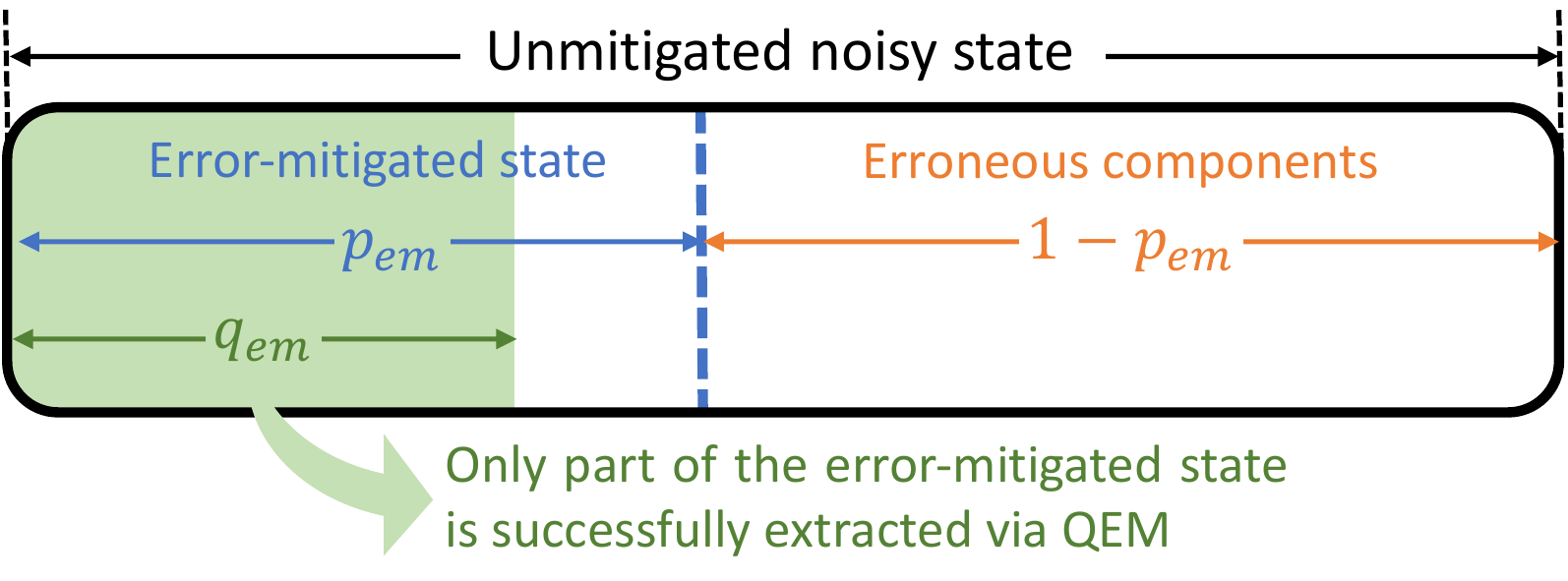}
    \caption{A diagram showing the different components within the original noisy state for a given QEM protocol.}
    \label{fig:state_extraction}
\end{figure}

\begin{figure*}[tb]
    \centering
    \includegraphics[width = 0.7\textwidth]{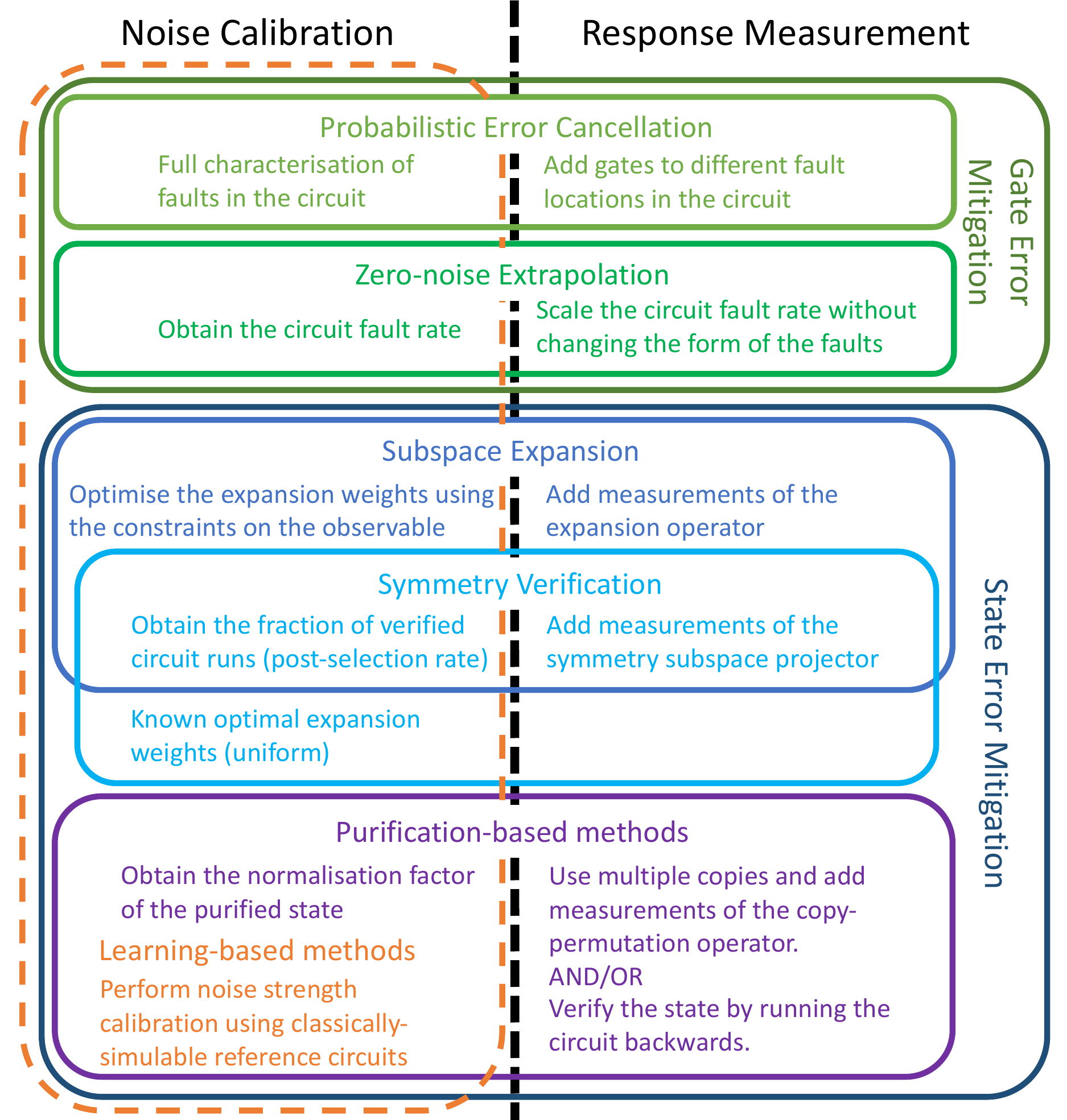}
    \caption{A diagram showing different categories of QEM methods and the structure within them.}
    \label{fig:venn_diagram}
\end{figure*}

In this article, we have decoupled the process of QEM into noise calibration and response measurement which enable us to systematically categorise different QEM methods as shown in \cref{fig:venn_diagram}. Then we introduced the framework of linear QEM, within which QEM can be viewed as extracting an error-mitigated state $\rho_{\mathrm{em}}$ out of the noisy state $\rho_{\lambda}$ as shown in \cref{fig:state_extraction}. The error-mitigated state $\rho_{\mathrm{em}}$ is the linear combination of the output `state' from a set of response measurement circuits, and their weightings are obtained through noise calibration. The amount of $\rho_{\mathrm{em}}$ contained within the unmitigated state $\rho_{\lambda}$, denoted as $p_{\mathrm{em}}$, will determine the amount of fidelity boost achieved by the QEM method over the unmitigated state. The amount of error-mitigated state that we can successfully extract via our QEM method, denoted as $q_{\mathrm{em}}$, will determine the sampling overhead of the method. The fraction of error-mitigated state that is successfully extracted $q_{\mathrm{em}}/p_{\mathrm{em}}$, which we call extraction rate, is the ratio between the fidelity boost and square root of sampling overhead. Thus, it serves as an indicator of the cost-effectiveness of a given QEM method. 

Under the practical assumptions of Markovian noise and orthogonal errors (\cref{eqn:ortho_error}), we have explicitly derived the fidelity boost, sampling cost and extraction rate of various QEM schemes. We found that symmetry verification can achieve the highest extraction rate possible of $1$, but the amount of noise it can remove is fixed and thus cannot be the only strategy that we employed in practice. The rest of the QEM schemes can remove a range of errors up to all errors in the circuit. Probabilistic error cancellation has the highest extraction rate among them, but the full noise characterisation required is often not fully available in practice. Comparing analytical extrapolation against multi-copy purification, we found that multi-copy purification has a higher fidelity boost and a higher extraction rate, when the number of copies is the same as the number of data points in analytical extrapolation. However, this does not mean that multi-copy purification is always preferred since it will require more copies of quantum devices, which can be used to run experiments in parallel in the other schemes. 

We see that sampling overhead is not the only resource cost that matters. A more illustrative metric for the total overhead could be the product of sampling overhead, qubit overhead and circuit runtime overhead. Qubit overhead is an essential component for symmetry verification. Since symmetry verification can be viewed as a quantum error detection (QED) code~\cite{mccleanDecodingQuantumErrors2020}, we may be able to connect the concepts in QEM to QED through our framework if we can understand the role of qubit overhead better, which in turn might connect to concepts in QEC. Such connections would be an exciting direction to explore. It is worth remembering that we have shown symmetry verification and more generally QED can extract all error-mitigated components contained in the noisy state (an extraction rate of $1$).

We have not talked about the extrapolation strategies with more data points than the number of parameters we try to fit~\cite{endoPracticalQuantumError2018,caiMultiexponentialErrorExtrapolation2021}. They will be an example of non-linear QEM estimators, within which we cannot view the process of quantum error mitigation as obtaining some error-mitigated quantum state anymore. They can only be viewed as performing post-processing directly on the measurement outputs instead of the underlying states. As we are no longer dealing with a quantum state, fidelity is not a relevant metric for non-linear estimators. Hence, one can further look into how to predict the performance of non-linear estimators analytically. 

Throughout this article, we have made a series of practical assumptions like pure ideal state, Markovian errors and orthogonal errors. Most of our arguments and order-of-magnitude estimates of various metrics will still be valid even if these assumptions are not entirely fulfilled. However, if we are looking at a situation far from these settings, further generalisation of our framework will be needed. This can also involve developing other metrics under our framework or merging some of the arguments in Ref.~\cite{takagiFundamentalLimitationsQuantum2021} into our framework.

\section*{Acknowledgements}
The author would like to thank Simon Benjamin for valuable discussions.

The author is supported by the Junior Research Fellowship from St John’s College, Oxford and acknowledges support from the QCS Hub (EP/T001062/1).

\appendix

\section{Obtaining Sampling Overhead using Hoeffding's inequality}\label{sec:samp_hoeff}
Here we define the \emph{range} of a random variable $\est{X}$ to be the difference between its maximal and minimal possible values, and we denote it as $\range{\est{X}}$. Since $\est{O}_{\mathrm{rsp}}$ is often obtained through measuring products of $O$ and operators that commute with $O$ and have eigenvalues bounded between $\pm 1$ (which include $O$ itself), we know that $\range{\est{O}_{\mathrm{rsp}}}\sim \range{\est{O}_{\rho_{\lambda}}}$. 

Hoeffding's inequality states that to estimate $\expect{\est{X}}$ to $\epsilon$-precision with $1-\delta$ probability, the number of samples we need is given by:
\begin{align*}
    N_{\mathrm{cir}}(\est{X}) = \frac{\ln(2/\delta)}{2\epsilon^2} \range{\est{X}}^2.
\end{align*}
Hence, for the additional number of circuit runs we need to estimate $\est{O}_{\mathrm{em}}$ to the same precision as $\est{O}_{\rho_{\lambda}}$ is given by:
\begin{align*}
    C_{\mathrm{em}} = \frac{N_{\mathrm{cir}}(\est{O}_{\mathrm{em}})}{N_{\mathrm{cir}}(\est{O}_{\rho_{\lambda}})} = \frac{\range{\est{O}_{\mathrm{em}}}^2}{\range{\est{O}_{\rho_{\lambda}}}^2} =\frac{q_{\mathrm{em}}^{-2}\range{\est{O}_{\mathrm{rsp}}}^2}{\range{\est{O}_{\rho_{\lambda}}}^2} \sim q_{\mathrm{em}}^{-2}. 
\end{align*}
where we have used \cref{eqn:linear_estimator}.

\section{Analytical Extrapolation}\label{sec:ana_extra}
\subsection{Error-mitigated State for Analytical Extrapolation}\label{sec:ana_extra_state}
Using \cref{eqn:noisy_state_distri,eqn:zne_em_state} the error-mitigated state of analytical extrapolation using $n$ data points can be rewritten as:
\begin{align}\label{eqn:zne_em_state_expand}
    \rho_{\mathrm{em}} &= \frac{1}{A}  \sum_{\ell = 0}^{\infty}  c_{\ell}(\vec{\lambda}) \frac{\lambda^\ell}{\ell!}\rho_{\abs{\mathbb{L}} = \ell}
\end{align}
where 
\begin{align}\label{eqn:c_form}
    c_{\ell}(\vec{\lambda}) = \sum_{i = 1}^{n} \gamma_i \left(\frac{\lambda_i}{\lambda}\right)^\ell.
\end{align}
Using Holder's inequality, we have
\begin{align*}
    c_{\ell}(\vec{\lambda}) \leq \left(\frac{\lambda_n}{\lambda}\right)^\ell \sum_{i = 1}^{n} \abs{\gamma_i}.
\end{align*}
Thus we have
\begin{align*}
    \rho_{\mathrm{em}} &\leq \frac{1}{A}  \sum_{\ell = 0}^{\infty}  \left(\sum_{i = 1}^{n} \abs{\gamma_i}\right) \frac{\lambda_n^\ell}{\ell!}\rho_{\abs{\mathbb{L}} = \ell}.
\end{align*}
which means that the sum in \cref{eqn:zne_em_state_expand} will converge as long as $\lambda_n$ and $\{\abs{\gamma_i}\}$ are finite, which implies $\abs{\lambda_i - \lambda_j}$ does not tends to zero or infinity for any $i,j$. 

From \cref{eqn:c_form}, we have $c_{0}(\vec{\lambda}) = 1$ and $c_{\ell}(\vec{\lambda}) = 0$  for $1 \leq \ell < n$ using the property of Lagrange polynomial. Thus, we have
\begin{align}
    \rho_{\mathrm{em}} &= \frac{1}{A}  \left(\rho_{0} +  \sum_{\ell = n}^{\infty}  c_{\ell}(\vec{\lambda}) \frac{\lambda^\ell}{\ell!}\rho_{\abs{\mathbb{L}} = \ell}\right).
\end{align}

\subsection{Odd Number of Data Points}\label{sec:odd_data_point}
Recall that 
\begin{align*}
    A = \sum_{j = 1}^{n} \gamma_j e^{\lambda_j}, \quad \sum_{j = 1}^{n} \gamma_j = 1.
\end{align*}
Since $\gamma_j = \prod_{k \neq j} \frac{\lambda_k}{\lambda_k - \lambda_j}$, $\gamma_j$ will be negative when there is an odd number of $\lambda_k$ smaller than $\lambda_j$ and will be positive otherwise. Since we have $\lambda_{j+1} >\lambda_j \quad \forall j$, $\gamma_j$ is positive for odd $j$ and negative for even $j$. Hence, we have:
\begin{align*}
    1 = \sum_{j = 1}^{n} \gamma_j = \begin{cases}
        \sum_{k = 0}^{n/2}  \abs{\gamma_{2k+1}} - \abs{\gamma_{2k+2}}\quad &\text{even }n\\
        \abs{\gamma_{1}} + \sum_{k = 1}^{(n+1)/2}  \abs{\gamma_{2k+1}} - \abs{\gamma_{2k}} \quad &\text{odd }n
    \end{cases}
\end{align*}
Now look at
\begin{align*}
    A &= \sum_{j = 1}^{n} \gamma_j e^{\lambda_j} \\
    &= \begin{cases}
        \sum_{k = 0}^{n/2}  \abs{\gamma_{2k+1}}e^{\lambda_{2k+1}} - \abs{\gamma_{2k+2}}e^{\lambda_{2k+2}}\quad &\text{even }n\\
        \abs{\gamma_{1}} + \sum_{k = 1}^{(n+1)/2}  \abs{\gamma_{2k+1}}e^{\lambda_{2k+1}} - \abs{\gamma_{2k}}e^{\lambda_{2k}} \quad &\text{odd }n
    \end{cases}
\end{align*}
Since $\lambda_{2k+1} < \lambda_{2k+2}$, we have
\begin{align*}
    e^{\lambda_{2k+1}} &< e^{\lambda_{2k+2}}\\
    \abs{\gamma_{2k+1}}e^{\lambda_{2k+1}} - \abs{\gamma_{2k+2}}e^{\lambda_{2k+2}} &< \abs{\gamma_{2k+1}} - \abs{\gamma_{2k+2}}\\
    A &< \sum_{j = 1}^{n} \gamma_j = 1 \quad \text{for even }n.
\end{align*}
Similarly, using $\lambda_{2k+1} > \lambda_{2k}$ we have
\begin{align*}
    A &> \sum_{j = 1}^{n} \gamma_j = 1 \quad \text{for odd }n.
\end{align*}
From \cref{eqn:zne_em_fid}, we see that the error-mitigated fidelity $\Tr(\rho_0\rho_{\mathrm{em}}) = A^{-1}$, thus we must have $A \geq 1$ otherwise our framework does not apply any more. This means that analytical extrapolation will only be applicable with odd number of data point (odd $n$). Hence, through out this article, we will assume we have odd number of data point (odd $n$). 

\subsection{Equal-gap Analytical Extrapolation}\label{sec:ana_extra_coef}

Suppose we are probing at error rates with a fixed gap $\beta$ and we further assume that the starting error rate $\lambda$ is an integer multiple of $\beta$, we then have
\begin{align*}
    \lambda_m = \lambda + (m-1)\beta = \left(m_0 + m-1\right)\beta
\end{align*}
We can then try to obtain an expression for the coefficient in Richardson extrapolation:
\begin{align*}
    \gamma_j &= \prod_{k \neq j} \frac{\lambda_k}{\lambda_k - \lambda_j} = \prod_{k \neq j}\frac{m_0 + k - 1}{k-j}\\
    & = \frac{n + \left(m_0-1\right)}{n-j} \times \cdots \times \frac{(j+1) + \left(m_0-1\right)}{(j+1)-j}\\
    & \times \frac{(j-1) + \left(m_0-1\right)}{(j-1)-j} \times \cdots \times\frac{1 + \left(m_0-1\right)}{1-j}\\
    & = \left(-1\right)^{j-1}\frac{\left(n + m_0-1\right)!}{\left(j + m_0-1\right) \left(m_0-1\right)! \left(n-j\right)!(j-1)!}\\
    & = \left(-1\right)^{j-1}\frac{j}{j + m_0-1} \binom{n + m_0-1}{n} \binom{n}{j}\\
    & = \left(-1\right)^{j-1}\frac{n + m_0-1}{j + m_0-1} \binom{n + m_0-2}{n-1} \binom{n-1}{j-1}
\end{align*}
Hence, we have
\begin{align}\label{eqn:B_equal_gap}
    &\quad A = \sum_{j = 1}^{n} \gamma_j e^{\lambda_j} \\
    &= \binom{n + m_0-1}{n}e^{m_0\beta} \sum_{j = 1}^{n}  \frac{j}{j + m_0-1}\binom{n}{j} \left(-e^{\beta}\right)^{j-1}\nonumber\\
    &= \binom{n + m_0-2}{n-1}e^{m_0\beta} \sum_{j = 1}^{n}  \frac{n + m_0-1}{j + m_0-1}\binom{n-1}{j-1} \left(-e^{\beta}\right)^{j-1}\nonumber
\end{align}
\begin{align}\label{eqn:A_equal_gap}
    &\quad A_{\mathrm{abs}} = \sum_{j = 1}^{n} \abs{\gamma_j} e^{\lambda_j} \\
    &= \binom{n + m_0-1}{n}e^{m_0\beta} \sum_{j = 1}^{n}\frac{j}{j + m_0-1}\binom{n}{j} e^{\left(j-1\right)\beta} \nonumber\\
    &= \binom{n + m_0-2}{n-1}e^{m_0\beta} \sum_{j = 1}^{n}\frac{n + m_0-1}{j + m_0-1}\binom{n-1}{j-1} e^{\left(j-1\right)\beta} \nonumber
\end{align}

Using \cref{eqn:A_equal_gap}, we can obtain a general lower bound for $A_{\mathrm{abs}}$:
\begin{align*}
    A_{\mathrm{abs}} &\geq \binom{n + m_0-2}{n-1}e^{m_0\beta} \sum_{j = 1}^{n} \binom{n-1}{j-1} e^{\left(j-1\right)\beta}\\
    & = \binom{n + m_0-2}{n-1}e^{m_0\beta} \left(1 + e^{\beta}\right)^{n-1}
\end{align*}
This can give a general upper bound on the extraction rate using \cref{eqn:zne_extraction_rate}:
\begin{align*}
    r_{\mathrm{em}} &= \frac{e^{\lambda}}{A_{\mathrm{abs}}} \leq  \binom{n + m_0-2}{n-1}^{-1}\left(1 + e^{\lambda/m_0}\right)^{1-n}
\end{align*}

At the practical value of $\lambda \sim 1$ and $n \sim 3$, we can explicitly calculate the extraction rate $r_{\mathrm{em}}$ and see that it peaks at $m_0 \sim 1$. Hence, here we will look at the case in which $m_0 = 1$, which means that $\lambda = \beta$ and $\lambda_n = n \lambda$. In this way we have:
\begin{align*}
    A &= - \sum_{j = 1}^{n}\binom{n}{j} \left(-e^{\beta}\right)^{j} =  1 - \left(1 - e^{\lambda}\right)^n\\
    A_{\mathrm{abs}} &=  \sum_{j = 1}^{n}\binom{n}{j} e^{j\beta} =  \left(1 + e^{\lambda}\right)^n - 1
\end{align*}
From \cref{sec:odd_data_point}, we know that we are only interested in odd $n$. Since $1 - e^{\lambda} \leq 0$, we have
\begin{align*}
    A & =  \left(e^{\lambda} - 1\right)^n + 1.
\end{align*}

Using \cref{eqn:zne_fid_boost,eqn:zne_samp_cost,eqn:zne_extraction_rate}, we can obtain the fidelity boost, sampling overhead and the extraction rate to be:
\begin{align*}
    B_{\mathrm{em}} &= \frac{e^{\lambda}}{A} = \frac{e^{\lambda}}{\left(e^{\lambda} - 1\right)^n + 1}\\
    C_{\mathrm{em}} &\sim \left(\frac{A_{\mathrm{abs}}}{A}\right)^2 = \left(\frac{\left(e^{\lambda}+1\right)^n - 1}{\left(e^{\lambda} - 1\right)^n + 1}\right)^2\\
    r_{\mathrm{em}} &= \frac{e^{\lambda}}{A_{\mathrm{abs}}} = \frac{e^{\lambda}}{\left(e^{\lambda}+1\right)^n - 1}\\
\end{align*}

\section{State Error Mitigation Sampling Overhead}\label{sec:state_samp_cost}
\begin{figure}[tb]
    \centering
    \includegraphics[width = 0.2\textwidth]{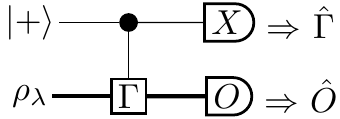}
    \caption{Hadamard Test Circuit. The expectation value of measuring $X\otimes O$, which is $\expect{\est{O}_{} \est{\Gamma}}$,  will be $\Re(\Tr(O\Gamma\rho))$. The expectation value of measuring $X\otimes I$, which is $\expect{\est{\Gamma}}$, will be $\Re(\Tr(\Gamma\rho))$}
    \label{fig:Htest}
\end{figure}
In post-processing symmetry verification~\cite{mcardleErrorMitigatedDigitalQuantum2019,bonet-monroigLowcostErrorMitigation2018} and multi-copy purification~\cite{koczorExponentialErrorSuppression2021,hugginsVirtualDistillationQuantum2021}, we need to measure two observables at the end of the noisy primary circuit using \cref{fig:Htest}, whose measurement outcomes corresponds to the random variables $\est{O}_{}$ and $\est{\Gamma}$, respectively. In this way, we can write the expectation value of the error-mitigated estimator as
\begin{align}\label{eqn:qem_expect}
    \expect{\est{O}_{\mathrm{em}}} = \frac{\expect{\est{O}_{} \est{\Gamma}}}{\expect{\est{\Gamma}}}.
\end{align}
In the $i$th circuit run, we will perform measurements to obtain $\est{O}_{\lambda,i}$ and $\est{\Gamma}_i$. After $N_{\mathrm{cir}}$ circuit runs, we can estimate $\expect{\est{O}_{\mathrm{em}}}$ through the estimator:
\begin{align}\label{eqn:mean_estimator}
    \mean{O}_{\mathrm{em}} = \frac{\sum_{i=1}^{N_{\mathrm{cir}}}\est{O}_{\lambda,i} \est{\Gamma}_i}{\sum_{i=1}^{N_{\mathrm{cir}}} \est{\Gamma}_i} = \frac{\mean{O_{}\Gamma}}{\mean{\Gamma}}.
\end{align}

Note that we have 
\begin{align*}
    \cov{\mean{O_{}\Gamma},\mean{\Gamma}} &= \frac{\sum_{i,j} \cov{\est{O}_{\lambda,i} \est{\Gamma}_i,\est{\Gamma}_j}}{N_{\mathrm{cir}}^2} = \frac{\sum_{i} \cov{\est{O}_{\lambda,i} \est{\Gamma}_i,\est{\Gamma}_i}}{N_{\mathrm{cir}}^2}\\ &= \frac{\cov{\est{O}_{} \est{\Gamma},\est{\Gamma}}}{N_{\mathrm{cir}}}
\end{align*}
Here we have used the fact that the measurements from different circuit runs are uncorrelated $\cov{\est{O}_{\lambda,i} \est{\Gamma}_i,\est{\Gamma}_j} = 0\quad \forall i\neq j$. 

Hence, the variance of $\mean{O}_{\mathrm{em}}$ is approximately

\begin{equation}
    \begin{aligned}\label{eqn:var_quotient_1}
        &\fiteqn{\var{\mean{O}_{\mathrm{em}}} = \frac{\var{\mean{O_{}\Gamma}} - 2\expect{\mean{O}_{\mathrm{em}}} \cov{\mean{O_{}\Gamma},\mean{\Gamma}} + \expect{\mean{O}_{\mathrm{em}}}^2 \var{\mean{\Gamma}}}{\expect{\est{\Gamma}}^2}} \\
        &= \frac{\var{\est{O}_{} \est{\Gamma}} - 2\expect{\est{O}_{\mathrm{em}}} \cov{\est{O}_{} \est{\Gamma},\est{\Gamma}} + \expect{\est{O}_{\mathrm{em}}}^2 \var{\est{\Gamma}}}{N_{\mathrm{cir}}\expect{\est{\Gamma}}^2} 
    \end{aligned}
\end{equation}
Assuming our observable $O$ is Pauli, which is the case for most of the practical settings. This means that $\est{O}_{} = \pm 1$ and thus  $\est{O}_{}^2 = 1$. Hence, we have:
\begin{align*}
    &\fiteqn{\var{\est{O}_{} \est{\Gamma}} = \expect{(\est{O}_{} \est{\Gamma})^2} - \expect{\est{O}_{} \est{\Gamma}}^2 = \expect{\est{\Gamma}^2} - \expect{\est{O}_{\mathrm{em}}}^2\expect{\est{\Gamma}}^2}\\
    &\var{\est{\Gamma}} = \expect{\est{\Gamma}^2} - \expect{\est{\Gamma}}^2\\
    &\fiteqn{\cov{\est{O}_{} \est{\Gamma}, \est{\Gamma}} = \expect{\est{O}_{} \est{\Gamma}^2} - \expect{\est{O}_{} \est{\Gamma}}\expect{\est{\Gamma}} = \expect{\est{O}_{} \est{\Gamma}^2} - \expect{\est{O}_{\mathrm{em}}}\expect{\est{\Gamma}}^2}
\end{align*} 
Substituting into \cref{eqn:var_quotient_1}, we have
\begin{equation}
    \begin{aligned}\label{eqn:var_quotient_2}
        \var{\mean{O}_{\mathrm{em}}} & = \frac{1}{N_{\mathrm{cir}}\expect{\est{\Gamma}}^2} \big[\expect{\est{\Gamma}^2} - \expect{\est{O}_{\mathrm{em}}}^2\expect{\est{\Gamma}}^2\\ 
        & \quad - 2\expect{\est{O}_{\mathrm{em}}}\expect{\est{O}_{} \est{\Gamma}^2} + 2\expect{\est{O}_{\mathrm{em}}}^2\expect{\est{\Gamma}}^2 \\
        & \quad + \expect{\est{O}_{\mathrm{em}}}^2\expect{\est{\Gamma}^2} -\expect{\est{O}_{\mathrm{em}}}^2 \expect{\est{\Gamma}}^2\big]\\
        & = \frac{\expect{\est{\Gamma}^2} - 2\expect{\est{O}_{\mathrm{em}}}\expect{\est{O}_{} \est{\Gamma}^2}+ \expect{\est{O}_{\mathrm{em}}}^2\expect{\est{\Gamma}^2}}{N_{\mathrm{cir}}\expect{\est{\Gamma}}^2}
    \end{aligned}
\end{equation}

Since $\est{\Gamma}$ is obtained through a Pauli measurement as shown in \cref{fig:Htest}, we have $\est{\Gamma} = \pm 1$ and we have $\est{\Gamma}^2 = 1$. Hence, \cref{eqn:var_quotient_2} becomes:
\begin{align*}
    \var{\mean{O}_{\mathrm{em}}} & = \frac{1 - 2\expect{\est{O}_{\mathrm{em}}}\expect{\est{O}_{}}+ \expect{\est{O}_{\mathrm{em}}}^2}{N_{\mathrm{cir}}\expect{\est{\Gamma}}^2}\\
    &\sim \frac{1 - \expect{\est{O}_{\rho_\lambda}}^2}{N_{\mathrm{cir}}\expect{\est{\Gamma}}^2} = \frac{\var{\mean{O}_{\rho_\lambda}}}{\expect{\est{\Gamma}}^2}.
\end{align*}
where we have assume $\expect{\est{O}_{\mathrm{em}}}  \sim \expect{\est{O}_{}} \sim \est{O}_{\rho_\lambda}$.

Hence, the sampling overhead is:
\begin{align*}
    C_{\mathrm{em}} = \frac{\var{\mean{O}_{\mathrm{em}}} }{\var{\mean{O}_{}}} \sim \expect{\est{\Gamma}}^{-2}
\end{align*}
which is what we have in \cref{eqn:sym_samp_cost,eqn:pur_samp_cost}.

\subsection{Direct Symmetry Verification}
In the case of direct symmetry verification, $\est{\Gamma}$ will be the measurement outcome of the projection operator $\Pi_{\mathbb{S}}$ instead of a Pauli operator. Hence $\est{\Gamma} = 0,1$ and we have $\est{\Gamma}^2 = \est{\Gamma}$. Hence, \cref{eqn:var_quotient_2} becomes:
\begin{align*}
    \var{\mean{O}_{\mathrm{em}}} & = \frac{\expect{\est{\Gamma}} - 2\expect{\est{O}_{\mathrm{em}}}\expect{\est{O}_{} \est{\Gamma}}+ \expect{\est{O}_{\mathrm{em}}}^2\expect{\est{\Gamma}}}{N_{\mathrm{cir}}\expect{\est{\Gamma}}^2}\\
    &= \frac{1 - \expect{\est{O}_{\mathrm{em}}}^2}{N_{\mathrm{cir}}\expect{\est{\Gamma}}}\\
    &\sim\frac{1 - \expect{\est{O}_{\rho_\lambda}}^2}{N_{\mathrm{cir}}\expect{\est{\Gamma}}} =  \frac{\var{\mean{O}_{\rho_\lambda}}}{\expect{\est{\Gamma}}}
\end{align*}
where we have use $\expect{\est{O}_{\mathrm{em}}} \sim \expect{\mean{O}_{\rho_\lambda}}$.

Hence, the sampling overhead is:
\begin{align*}
    C_{\mathrm{em}} = \frac{\var{\mean{O}_{\mathrm{em}}} }{\var{\mean{O}_{}}} \sim \expect{\est{\Gamma}}^{-1}.
\end{align*}

%\bibliographystyle{natbib}
% \bibliography{ref}

\begin{thebibliography}{36}%
\makeatletter
\providecommand \@ifxundefined [1]{%
 \@ifx{#1\undefined}
}%
\providecommand \@ifnum [1]{%
 \ifnum #1\expandafter \@firstoftwo
 \else \expandafter \@secondoftwo
 \fi
}%
\providecommand \@ifx [1]{%
 \ifx #1\expandafter \@firstoftwo
 \else \expandafter \@secondoftwo
 \fi
}%
\providecommand \natexlab [1]{#1}%
\providecommand \enquote  [1]{``#1''}%
\providecommand \bibnamefont  [1]{#1}%
\providecommand \bibfnamefont [1]{#1}%
\providecommand \citenamefont [1]{#1}%
\providecommand \href@noop [0]{\@secondoftwo}%
\providecommand \href [0]{\begingroup \@sanitize@url \@href}%
\providecommand \@href[1]{\@@startlink{#1}\@@href}%
\providecommand \@@href[1]{\endgroup#1\@@endlink}%
\providecommand \@sanitize@url [0]{\catcode `\\12\catcode `\$12\catcode
  `\&12\catcode `\#12\catcode `\^12\catcode `\_12\catcode `\%12\relax}%
\providecommand \@@startlink[1]{}%
\providecommand \@@endlink[0]{}%
\providecommand \url  [0]{\begingroup\@sanitize@url \@url }%
\providecommand \@url [1]{\endgroup\@href {#1}{\urlprefix }}%
\providecommand \urlprefix  [0]{URL }%
\providecommand \Eprint [0]{\href }%
\providecommand \doibase [0]{https://doi.org/}%
\providecommand \selectlanguage [0]{\@gobble}%
\providecommand \bibinfo  [0]{\@secondoftwo}%
\providecommand \bibfield  [0]{\@secondoftwo}%
\providecommand \translation [1]{[#1]}%
\providecommand \BibitemOpen [0]{}%
\providecommand \bibitemStop [0]{}%
\providecommand \bibitemNoStop [0]{.\EOS\space}%
\providecommand \EOS [0]{\spacefactor3000\relax}%
\providecommand \BibitemShut  [1]{\csname bibitem#1\endcsname}%
\let\auto@bib@innerbib\@empty
%</preamble>
\bibitem [{\citenamefont {Terhal}(2015)}]{terhalQuantumErrorCorrection2015}%
  \BibitemOpen
  \bibfield  {author} {\bibinfo {author} {\bibfnamefont {B.~M.}\ \bibnamefont
  {Terhal}},\ }\bibfield  {title} {\bibinfo {title} {Quantum error correction
  for quantum memories},\ }\href {https://doi.org/10.1103/RevModPhys.87.307}
  {\bibfield  {journal} {\bibinfo  {journal} {Reviews of Modern Physics}\
  }\textbf {\bibinfo {volume} {87}},\ \bibinfo {pages} {307} (\bibinfo {year}
  {2015})}\BibitemShut {NoStop}%
\bibitem [{\citenamefont {Endo}\ \emph {et~al.}(2021)\citenamefont {Endo},
  \citenamefont {Cai}, \citenamefont {Benjamin},\ and\ \citenamefont
  {Yuan}}]{endoHybridQuantumClassicalAlgorithms2021}%
  \BibitemOpen
  \bibfield  {author} {\bibinfo {author} {\bibfnamefont {S.}~\bibnamefont
  {Endo}}, \bibinfo {author} {\bibfnamefont {Z.}~\bibnamefont {Cai}}, \bibinfo
  {author} {\bibfnamefont {S.~C.}\ \bibnamefont {Benjamin}},\ and\ \bibinfo
  {author} {\bibfnamefont {X.}~\bibnamefont {Yuan}},\ }\bibfield  {title}
  {\bibinfo {title} {Hybrid {{Quantum}}-{{Classical Algorithms}} and {{Quantum
  Error Mitigation}}},\ }\href {https://doi.org/10.7566/JPSJ.90.032001}
  {\bibfield  {journal} {\bibinfo  {journal} {Journal of the Physical Society
  of Japan}\ }\textbf {\bibinfo {volume} {90}},\ \bibinfo {pages} {032001}
  (\bibinfo {year} {2021})}\BibitemShut {NoStop}%
\bibitem [{\citenamefont {Kandala}\ \emph {et~al.}(2019)\citenamefont
  {Kandala}, \citenamefont {Temme}, \citenamefont {C{\'o}rcoles}, \citenamefont
  {Mezzacapo}, \citenamefont {Chow},\ and\ \citenamefont
  {Gambetta}}]{kandalaErrorMitigationExtends2019}%
  \BibitemOpen
  \bibfield  {author} {\bibinfo {author} {\bibfnamefont {A.}~\bibnamefont
  {Kandala}}, \bibinfo {author} {\bibfnamefont {K.}~\bibnamefont {Temme}},
  \bibinfo {author} {\bibfnamefont {A.~D.}\ \bibnamefont {C{\'o}rcoles}},
  \bibinfo {author} {\bibfnamefont {A.}~\bibnamefont {Mezzacapo}}, \bibinfo
  {author} {\bibfnamefont {J.~M.}\ \bibnamefont {Chow}},\ and\ \bibinfo
  {author} {\bibfnamefont {J.~M.}\ \bibnamefont {Gambetta}},\ }\bibfield
  {title} {\bibinfo {title} {Error mitigation extends the computational reach
  of a noisy quantum processor},\ }\href
  {https://doi.org/10.1038/s41586-019-1040-7} {\bibfield  {journal} {\bibinfo
  {journal} {Nature}\ }\textbf {\bibinfo {volume} {567}},\ \bibinfo {pages}
  {491} (\bibinfo {year} {2019})}\BibitemShut {NoStop}%
\bibitem [{\citenamefont {Sagastizabal}\ \emph {et~al.}(2019)\citenamefont
  {Sagastizabal}, \citenamefont {{Bonet-Monroig}}, \citenamefont {Singh},
  \citenamefont {Rol}, \citenamefont {Bultink}, \citenamefont {Fu},
  \citenamefont {Price}, \citenamefont {Ostroukh}, \citenamefont
  {Muthusubramanian}, \citenamefont {Bruno}, \citenamefont {Beekman},
  \citenamefont {Haider}, \citenamefont {O'Brien},\ and\ \citenamefont
  {DiCarlo}}]{sagastizabalExperimentalErrorMitigation2019}%
  \BibitemOpen
  \bibfield  {author} {\bibinfo {author} {\bibfnamefont {R.}~\bibnamefont
  {Sagastizabal}}, \bibinfo {author} {\bibfnamefont {X.}~\bibnamefont
  {{Bonet-Monroig}}}, \bibinfo {author} {\bibfnamefont {M.}~\bibnamefont
  {Singh}}, \bibinfo {author} {\bibfnamefont {M.~A.}\ \bibnamefont {Rol}},
  \bibinfo {author} {\bibfnamefont {C.~C.}\ \bibnamefont {Bultink}}, \bibinfo
  {author} {\bibfnamefont {X.}~\bibnamefont {Fu}}, \bibinfo {author}
  {\bibfnamefont {C.~H.}\ \bibnamefont {Price}}, \bibinfo {author}
  {\bibfnamefont {V.~P.}\ \bibnamefont {Ostroukh}}, \bibinfo {author}
  {\bibfnamefont {N.}~\bibnamefont {Muthusubramanian}}, \bibinfo {author}
  {\bibfnamefont {A.}~\bibnamefont {Bruno}}, \bibinfo {author} {\bibfnamefont
  {M.}~\bibnamefont {Beekman}}, \bibinfo {author} {\bibfnamefont
  {N.}~\bibnamefont {Haider}}, \bibinfo {author} {\bibfnamefont {T.~E.}\
  \bibnamefont {O'Brien}},\ and\ \bibinfo {author} {\bibfnamefont
  {L.}~\bibnamefont {DiCarlo}},\ }\bibfield  {title} {\bibinfo {title}
  {Experimental error mitigation via symmetry verification in a variational
  quantum eigensolver},\ }\href {https://doi.org/10.1103/PhysRevA.100.010302}
  {\bibfield  {journal} {\bibinfo  {journal} {Physical Review A}\ }\textbf
  {\bibinfo {volume} {100}},\ \bibinfo {pages} {010302} (\bibinfo {year}
  {2019})}\BibitemShut {NoStop}%
\bibitem [{\citenamefont {Song}\ \emph {et~al.}(2019)\citenamefont {Song},
  \citenamefont {Cui}, \citenamefont {Wang}, \citenamefont {Hao}, \citenamefont
  {Feng},\ and\ \citenamefont {Li}}]{songQuantumComputationUniversal2019}%
  \BibitemOpen
  \bibfield  {author} {\bibinfo {author} {\bibfnamefont {C.}~\bibnamefont
  {Song}}, \bibinfo {author} {\bibfnamefont {J.}~\bibnamefont {Cui}}, \bibinfo
  {author} {\bibfnamefont {H.}~\bibnamefont {Wang}}, \bibinfo {author}
  {\bibfnamefont {J.}~\bibnamefont {Hao}}, \bibinfo {author} {\bibfnamefont
  {H.}~\bibnamefont {Feng}},\ and\ \bibinfo {author} {\bibfnamefont
  {Y.}~\bibnamefont {Li}},\ }\bibfield  {title} {\bibinfo {title} {Quantum
  computation with universal error mitigation on a superconducting quantum
  processor},\ }\href {https://doi.org/10.1126/sciadv.aaw5686} {\bibfield
  {journal} {\bibinfo  {journal} {Science Advances}\ }\textbf {\bibinfo
  {volume} {5}},\ \bibinfo {pages} {eaaw5686} (\bibinfo {year}
  {2019})}\BibitemShut {NoStop}%
\bibitem [{\citenamefont {{Google AI Quantum {and}
  Collaborators}}(2020{\natexlab{a}})}]{googleaiquantumandcollaboratorsHartreeFockSuperconductingQubit2020}%
  \BibitemOpen
  \bibfield  {author} {\bibinfo {author} {\bibnamefont {{Google AI Quantum
  {and} Collaborators}}},\ }\bibfield  {title} {\bibinfo {title}
  {Hartree-{{Fock}} on a superconducting qubit quantum computer},\ }\href
  {https://doi.org/10.1126/science.abb9811} {\bibfield  {journal} {\bibinfo
  {journal} {Science}\ }\textbf {\bibinfo {volume} {369}},\ \bibinfo {pages}
  {1084} (\bibinfo {year} {2020}{\natexlab{a}})}\BibitemShut {NoStop}%
\bibitem [{\citenamefont {{Google AI Quantum {and}
  Collaborators}}(2020{\natexlab{b}})}]{googleaiquantumandcollaboratorsObservationSeparatedDynamics2020}%
  \BibitemOpen
  \bibfield  {author} {\bibinfo {author} {\bibnamefont {{Google AI Quantum
  {and} Collaborators}}},\ }\bibfield  {title} {\bibinfo {title} {Observation
  of separated dynamics of charge and spin in the {{Fermi}}-{{Hubbard}}
  model},\ }\href {http://arxiv.org/abs/2010.07965} {\bibfield  {journal}
  {\bibinfo  {journal} {arXiv:2010.07965 [quant-ph]}\ } (\bibinfo {year}
  {2020}{\natexlab{b}})}\BibitemShut {NoStop}%
\bibitem [{\citenamefont {Suzuki}\ \emph {et~al.}(2021)\citenamefont {Suzuki},
  \citenamefont {Endo}, \citenamefont {Fujii},\ and\ \citenamefont
  {Tokunaga}}]{suzukiQuantumErrorMitigation2021}%
  \BibitemOpen
  \bibfield  {author} {\bibinfo {author} {\bibfnamefont {Y.}~\bibnamefont
  {Suzuki}}, \bibinfo {author} {\bibfnamefont {S.}~\bibnamefont {Endo}},
  \bibinfo {author} {\bibfnamefont {K.}~\bibnamefont {Fujii}},\ and\ \bibinfo
  {author} {\bibfnamefont {Y.}~\bibnamefont {Tokunaga}},\ }\bibfield  {title}
  {\bibinfo {title} {Quantum error mitigation for fault-tolerant quantum
  computing},\ }\href {http://arxiv.org/abs/2010.03887} {\bibfield  {journal}
  {\bibinfo  {journal} {arXiv:2010.03887 [quant-ph]}\ } (\bibinfo {year}
  {2021})}\BibitemShut {NoStop}%
\bibitem [{\citenamefont {Piveteau}\ \emph {et~al.}(2021)\citenamefont
  {Piveteau}, \citenamefont {Sutter}, \citenamefont {Bravyi}, \citenamefont
  {Gambetta},\ and\ \citenamefont
  {Temme}}]{piveteauErrorMitigationUniversal2021}%
  \BibitemOpen
  \bibfield  {author} {\bibinfo {author} {\bibfnamefont {C.}~\bibnamefont
  {Piveteau}}, \bibinfo {author} {\bibfnamefont {D.}~\bibnamefont {Sutter}},
  \bibinfo {author} {\bibfnamefont {S.}~\bibnamefont {Bravyi}}, \bibinfo
  {author} {\bibfnamefont {J.~M.}\ \bibnamefont {Gambetta}},\ and\ \bibinfo
  {author} {\bibfnamefont {K.}~\bibnamefont {Temme}},\ }\bibfield  {title}
  {\bibinfo {title} {Error mitigation for universal gates on encoded qubits},\
  }\href {http://arxiv.org/abs/2103.04915} {\bibfield  {journal} {\bibinfo
  {journal} {arXiv:2103.04915 [quant-ph]}\ } (\bibinfo {year}
  {2021})}\BibitemShut {NoStop}%
\bibitem [{\citenamefont {Temme}\ \emph {et~al.}(2017)\citenamefont {Temme},
  \citenamefont {Bravyi},\ and\ \citenamefont
  {Gambetta}}]{temmeErrorMitigationShortDepth2017}%
  \BibitemOpen
  \bibfield  {author} {\bibinfo {author} {\bibfnamefont {K.}~\bibnamefont
  {Temme}}, \bibinfo {author} {\bibfnamefont {S.}~\bibnamefont {Bravyi}},\ and\
  \bibinfo {author} {\bibfnamefont {J.~M.}\ \bibnamefont {Gambetta}},\
  }\bibfield  {title} {\bibinfo {title} {Error {{Mitigation}} for
  {{Short}}-{{Depth Quantum Circuits}}},\ }\href
  {https://doi.org/10.1103/PhysRevLett.119.180509} {\bibfield  {journal}
  {\bibinfo  {journal} {Physical Review Letters}\ }\textbf {\bibinfo {volume}
  {119}},\ \bibinfo {pages} {180509} (\bibinfo {year} {2017})}\BibitemShut
  {NoStop}%
\bibitem [{\citenamefont {Endo}\ \emph {et~al.}(2018)\citenamefont {Endo},
  \citenamefont {Benjamin},\ and\ \citenamefont
  {Li}}]{endoPracticalQuantumError2018}%
  \BibitemOpen
  \bibfield  {author} {\bibinfo {author} {\bibfnamefont {S.}~\bibnamefont
  {Endo}}, \bibinfo {author} {\bibfnamefont {S.~C.}\ \bibnamefont {Benjamin}},\
  and\ \bibinfo {author} {\bibfnamefont {Y.}~\bibnamefont {Li}},\ }\bibfield
  {title} {\bibinfo {title} {Practical {{Quantum Error Mitigation}} for
  {{Near}}-{{Future Applications}}},\ }\href
  {https://doi.org/10.1103/PhysRevX.8.031027} {\bibfield  {journal} {\bibinfo
  {journal} {Physical Review X}\ }\textbf {\bibinfo {volume} {8}},\ \bibinfo
  {pages} {031027} (\bibinfo {year} {2018})}\BibitemShut {NoStop}%
\bibitem [{\citenamefont {Li}\ and\ \citenamefont
  {Benjamin}(2017)}]{liEfficientVariationalQuantum2017}%
  \BibitemOpen
  \bibfield  {author} {\bibinfo {author} {\bibfnamefont {Y.}~\bibnamefont
  {Li}}\ and\ \bibinfo {author} {\bibfnamefont {S.~C.}\ \bibnamefont
  {Benjamin}},\ }\bibfield  {title} {\bibinfo {title} {Efficient {{Variational
  Quantum Simulator Incorporating Active Error Minimization}}},\ }\href
  {https://doi.org/10.1103/PhysRevX.7.021050} {\bibfield  {journal} {\bibinfo
  {journal} {Physical Review X}\ }\textbf {\bibinfo {volume} {7}},\ \bibinfo
  {pages} {021050} (\bibinfo {year} {2017})}\BibitemShut {NoStop}%
\bibitem [{\citenamefont
  {Cai}(2021{\natexlab{a}})}]{caiMultiexponentialErrorExtrapolation2021}%
  \BibitemOpen
  \bibfield  {author} {\bibinfo {author} {\bibfnamefont {Z.}~\bibnamefont
  {Cai}},\ }\bibfield  {title} {\bibinfo {title} {Multi-exponential error
  extrapolation and combining error mitigation techniques for {{NISQ}}
  applications},\ }\href {https://doi.org/10.1038/s41534-021-00404-3}
  {\bibfield  {journal} {\bibinfo  {journal} {npj Quantum Information}\
  }\textbf {\bibinfo {volume} {7}},\ \bibinfo {pages} {1} (\bibinfo {year}
  {2021}{\natexlab{a}})}\BibitemShut {NoStop}%
\bibitem [{\citenamefont {McClean}\ \emph {et~al.}(2017)\citenamefont
  {McClean}, \citenamefont {{Kimchi-Schwartz}}, \citenamefont {Carter},\ and\
  \citenamefont {{de Jong}}}]{mccleanHybridQuantumclassicalHierarchy2017}%
  \BibitemOpen
  \bibfield  {author} {\bibinfo {author} {\bibfnamefont {J.~R.}\ \bibnamefont
  {McClean}}, \bibinfo {author} {\bibfnamefont {M.~E.}\ \bibnamefont
  {{Kimchi-Schwartz}}}, \bibinfo {author} {\bibfnamefont {J.}~\bibnamefont
  {Carter}},\ and\ \bibinfo {author} {\bibfnamefont {W.~A.}\ \bibnamefont {{de
  Jong}}},\ }\bibfield  {title} {\bibinfo {title} {Hybrid quantum-classical
  hierarchy for mitigation of decoherence and determination of excited
  states},\ }\href {https://doi.org/10.1103/PhysRevA.95.042308} {\bibfield
  {journal} {\bibinfo  {journal} {Physical Review A}\ }\textbf {\bibinfo
  {volume} {95}},\ \bibinfo {pages} {042308} (\bibinfo {year}
  {2017})}\BibitemShut {NoStop}%
\bibitem [{\citenamefont {McArdle}\ \emph {et~al.}(2019)\citenamefont
  {McArdle}, \citenamefont {Yuan},\ and\ \citenamefont
  {Benjamin}}]{mcardleErrorMitigatedDigitalQuantum2019}%
  \BibitemOpen
  \bibfield  {author} {\bibinfo {author} {\bibfnamefont {S.}~\bibnamefont
  {McArdle}}, \bibinfo {author} {\bibfnamefont {X.}~\bibnamefont {Yuan}},\ and\
  \bibinfo {author} {\bibfnamefont {S.}~\bibnamefont {Benjamin}},\ }\bibfield
  {title} {\bibinfo {title} {Error-{{Mitigated Digital Quantum Simulation}}},\
  }\href {https://doi.org/10.1103/PhysRevLett.122.180501} {\bibfield  {journal}
  {\bibinfo  {journal} {Physical Review Letters}\ }\textbf {\bibinfo {volume}
  {122}},\ \bibinfo {pages} {180501} (\bibinfo {year} {2019})}\BibitemShut
  {NoStop}%
\bibitem [{\citenamefont {{Bonet-Monroig}}\ \emph {et~al.}(2018)\citenamefont
  {{Bonet-Monroig}}, \citenamefont {Sagastizabal}, \citenamefont {Singh},\ and\
  \citenamefont {O'Brien}}]{bonet-monroigLowcostErrorMitigation2018}%
  \BibitemOpen
  \bibfield  {author} {\bibinfo {author} {\bibfnamefont {X.}~\bibnamefont
  {{Bonet-Monroig}}}, \bibinfo {author} {\bibfnamefont {R.}~\bibnamefont
  {Sagastizabal}}, \bibinfo {author} {\bibfnamefont {M.}~\bibnamefont
  {Singh}},\ and\ \bibinfo {author} {\bibfnamefont {T.~E.}\ \bibnamefont
  {O'Brien}},\ }\bibfield  {title} {\bibinfo {title} {Low-cost error mitigation
  by symmetry verification},\ }\href
  {https://doi.org/10.1103/PhysRevA.98.062339} {\bibfield  {journal} {\bibinfo
  {journal} {Physical Review A}\ }\textbf {\bibinfo {volume} {98}},\ \bibinfo
  {pages} {062339} (\bibinfo {year} {2018})}\BibitemShut {NoStop}%
\bibitem [{\citenamefont
  {Cai}(2021{\natexlab{b}})}]{caiQuantumErrorMitigation2021a}%
  \BibitemOpen
  \bibfield  {author} {\bibinfo {author} {\bibfnamefont {Z.}~\bibnamefont
  {Cai}},\ }\bibfield  {title} {\bibinfo {title} {Quantum {{Error Mitigation}}
  using {{Symmetry Expansion}}},\ }\href
  {https://doi.org/10.22331/q-2021-09-21-548} {\bibfield  {journal} {\bibinfo
  {journal} {Quantum}\ }\textbf {\bibinfo {volume} {5}},\ \bibinfo {pages}
  {548} (\bibinfo {year} {2021}{\natexlab{b}})}\BibitemShut {NoStop}%
\bibitem [{\citenamefont
  {Koczor}(2021{\natexlab{a}})}]{koczorExponentialErrorSuppression2021}%
  \BibitemOpen
  \bibfield  {author} {\bibinfo {author} {\bibfnamefont {B.}~\bibnamefont
  {Koczor}},\ }\bibfield  {title} {\bibinfo {title} {Exponential {{Error
  Suppression}} for {{Near}}-{{Term Quantum Devices}}},\ }\href
  {https://doi.org/10.1103/PhysRevX.11.031057} {\bibfield  {journal} {\bibinfo
  {journal} {Physical Review X}\ }\textbf {\bibinfo {volume} {11}},\ \bibinfo
  {pages} {031057} (\bibinfo {year} {2021}{\natexlab{a}})}\BibitemShut
  {NoStop}%
\bibitem [{\citenamefont {Huggins}\ \emph {et~al.}(2021)\citenamefont
  {Huggins}, \citenamefont {McArdle}, \citenamefont {O'Brien}, \citenamefont
  {Lee}, \citenamefont {Rubin}, \citenamefont {Boixo}, \citenamefont {Whaley},
  \citenamefont {Babbush},\ and\ \citenamefont
  {McClean}}]{hugginsVirtualDistillationQuantum2021}%
  \BibitemOpen
  \bibfield  {author} {\bibinfo {author} {\bibfnamefont {W.~J.}\ \bibnamefont
  {Huggins}}, \bibinfo {author} {\bibfnamefont {S.}~\bibnamefont {McArdle}},
  \bibinfo {author} {\bibfnamefont {T.~E.}\ \bibnamefont {O'Brien}}, \bibinfo
  {author} {\bibfnamefont {J.}~\bibnamefont {Lee}}, \bibinfo {author}
  {\bibfnamefont {N.~C.}\ \bibnamefont {Rubin}}, \bibinfo {author}
  {\bibfnamefont {S.}~\bibnamefont {Boixo}}, \bibinfo {author} {\bibfnamefont
  {K.~B.}\ \bibnamefont {Whaley}}, \bibinfo {author} {\bibfnamefont
  {R.}~\bibnamefont {Babbush}},\ and\ \bibinfo {author} {\bibfnamefont {J.~R.}\
  \bibnamefont {McClean}},\ }\bibfield  {title} {\bibinfo {title} {Virtual
  {{Distillation}} for {{Quantum Error Mitigation}}},\ }\href
  {http://arxiv.org/abs/2011.07064} {\bibfield  {journal} {\bibinfo  {journal}
  {arXiv:2011.07064 [quant-ph]}\ } (\bibinfo {year} {2021})}\BibitemShut
  {NoStop}%
\bibitem [{\citenamefont {Huo}\ and\ \citenamefont
  {Li}(2021)}]{huoDualstatePurificationPractical2021}%
  \BibitemOpen
  \bibfield  {author} {\bibinfo {author} {\bibfnamefont {M.}~\bibnamefont
  {Huo}}\ and\ \bibinfo {author} {\bibfnamefont {Y.}~\bibnamefont {Li}},\
  }\bibfield  {title} {\bibinfo {title} {Dual-state purification for practical
  quantum error mitigation},\ }\href {http://arxiv.org/abs/2105.01239}
  {\bibfield  {journal} {\bibinfo  {journal} {arXiv:2105.01239 [quant-ph]}\ }
  (\bibinfo {year} {2021})}\BibitemShut {NoStop}%
\bibitem [{\citenamefont
  {Cai}(2021{\natexlab{c}})}]{caiResourceefficientPurificationbasedQuantum2021}%
  \BibitemOpen
  \bibfield  {author} {\bibinfo {author} {\bibfnamefont {Z.}~\bibnamefont
  {Cai}},\ }\bibfield  {title} {\bibinfo {title} {Resource-efficient
  {{Purification}}-based {{Quantum Error Mitigation}}},\ }\href
  {http://arxiv.org/abs/2107.07279} {\bibfield  {journal} {\bibinfo  {journal}
  {arXiv:2107.07279 [quant-ph]}\ } (\bibinfo {year}
  {2021}{\natexlab{c}})}\BibitemShut {NoStop}%
\bibitem [{\citenamefont {Bultrini}\ \emph {et~al.}(2021)\citenamefont
  {Bultrini}, \citenamefont {Gordon}, \citenamefont {Czarnik}, \citenamefont
  {Arrasmith}, \citenamefont {Coles},\ and\ \citenamefont
  {Cincio}}]{bultriniUnifyingBenchmarkingStateoftheart2021}%
  \BibitemOpen
  \bibfield  {author} {\bibinfo {author} {\bibfnamefont {D.}~\bibnamefont
  {Bultrini}}, \bibinfo {author} {\bibfnamefont {M.~H.}\ \bibnamefont
  {Gordon}}, \bibinfo {author} {\bibfnamefont {P.}~\bibnamefont {Czarnik}},
  \bibinfo {author} {\bibfnamefont {A.}~\bibnamefont {Arrasmith}}, \bibinfo
  {author} {\bibfnamefont {P.~J.}\ \bibnamefont {Coles}},\ and\ \bibinfo
  {author} {\bibfnamefont {L.}~\bibnamefont {Cincio}},\ }\bibfield  {title}
  {\bibinfo {title} {Unifying and benchmarking state-of-the-art quantum error
  mitigation techniques},\ }\href {http://arxiv.org/abs/2107.13470} {\bibfield
  {journal} {\bibinfo  {journal} {arXiv:2107.13470 [quant-ph]}\ } (\bibinfo
  {year} {2021})}\BibitemShut {NoStop}%
\bibitem [{\citenamefont {Mari}\ \emph {et~al.}(2021)\citenamefont {Mari},
  \citenamefont {Shammah},\ and\ \citenamefont
  {Zeng}}]{mariExtendingQuantumProbabilistic2021}%
  \BibitemOpen
  \bibfield  {author} {\bibinfo {author} {\bibfnamefont {A.}~\bibnamefont
  {Mari}}, \bibinfo {author} {\bibfnamefont {N.}~\bibnamefont {Shammah}},\ and\
  \bibinfo {author} {\bibfnamefont {W.~J.}\ \bibnamefont {Zeng}},\ }\bibfield
  {title} {\bibinfo {title} {Extending quantum probabilistic error cancellation
  by noise scaling},\ }\href {http://arxiv.org/abs/2108.02237} {\bibfield
  {journal} {\bibinfo  {journal} {arXiv:2108.02237 [quant-ph]}\ } (\bibinfo
  {year} {2021})}\BibitemShut {NoStop}%
\bibitem [{\citenamefont {Lowe}\ \emph {et~al.}(2021)\citenamefont {Lowe},
  \citenamefont {Gordon}, \citenamefont {Czarnik}, \citenamefont {Arrasmith},
  \citenamefont {Coles},\ and\ \citenamefont
  {Cincio}}]{loweUnifiedApproachDatadriven2021}%
  \BibitemOpen
  \bibfield  {author} {\bibinfo {author} {\bibfnamefont {A.}~\bibnamefont
  {Lowe}}, \bibinfo {author} {\bibfnamefont {M.~H.}\ \bibnamefont {Gordon}},
  \bibinfo {author} {\bibfnamefont {P.}~\bibnamefont {Czarnik}}, \bibinfo
  {author} {\bibfnamefont {A.}~\bibnamefont {Arrasmith}}, \bibinfo {author}
  {\bibfnamefont {P.~J.}\ \bibnamefont {Coles}},\ and\ \bibinfo {author}
  {\bibfnamefont {L.}~\bibnamefont {Cincio}},\ }\bibfield  {title} {\bibinfo
  {title} {Unified approach to data-driven quantum error mitigation},\ }\href
  {https://doi.org/10.1103/PhysRevResearch.3.033098} {\bibfield  {journal}
  {\bibinfo  {journal} {Physical Review Research}\ }\textbf {\bibinfo {volume}
  {3}},\ \bibinfo {pages} {033098} (\bibinfo {year} {2021})}\BibitemShut
  {NoStop}%
\bibitem [{\citenamefont {Wang}\ \emph {et~al.}(2021)\citenamefont {Wang},
  \citenamefont {Czarnik}, \citenamefont {Arrasmith}, \citenamefont {Cerezo},
  \citenamefont {Cincio},\ and\ \citenamefont
  {Coles}}]{wangCanErrorMitigation2021}%
  \BibitemOpen
  \bibfield  {author} {\bibinfo {author} {\bibfnamefont {S.}~\bibnamefont
  {Wang}}, \bibinfo {author} {\bibfnamefont {P.}~\bibnamefont {Czarnik}},
  \bibinfo {author} {\bibfnamefont {A.}~\bibnamefont {Arrasmith}}, \bibinfo
  {author} {\bibfnamefont {M.}~\bibnamefont {Cerezo}}, \bibinfo {author}
  {\bibfnamefont {L.}~\bibnamefont {Cincio}},\ and\ \bibinfo {author}
  {\bibfnamefont {P.~J.}\ \bibnamefont {Coles}},\ }\bibfield  {title} {\bibinfo
  {title} {Can {{Error Mitigation Improve Trainability}} of {{Noisy Variational
  Quantum Algorithms}}?},\ }\href {http://arxiv.org/abs/2109.01051} {\bibfield
  {journal} {\bibinfo  {journal} {arXiv:2109.01051 [quant-ph]}\ } (\bibinfo
  {year} {2021})}\BibitemShut {NoStop}%
\bibitem [{\citenamefont {Takagi}\ \emph {et~al.}(2021)\citenamefont {Takagi},
  \citenamefont {Endo}, \citenamefont {Minagawa},\ and\ \citenamefont
  {Gu}}]{takagiFundamentalLimitationsQuantum2021}%
  \BibitemOpen
  \bibfield  {author} {\bibinfo {author} {\bibfnamefont {R.}~\bibnamefont
  {Takagi}}, \bibinfo {author} {\bibfnamefont {S.}~\bibnamefont {Endo}},
  \bibinfo {author} {\bibfnamefont {S.}~\bibnamefont {Minagawa}},\ and\
  \bibinfo {author} {\bibfnamefont {M.}~\bibnamefont {Gu}},\ }\bibfield
  {title} {\bibinfo {title} {Fundamental limitations of quantum error
  mitigation},\ }\href {http://arxiv.org/abs/2109.04457} {\bibfield  {journal}
  {\bibinfo  {journal} {arXiv:2109.04457 [quant-ph]}\ } (\bibinfo {year}
  {2021})}\BibitemShut {NoStop}%
\bibitem [{\citenamefont
  {Koczor}(2021{\natexlab{b}})}]{koczorDominantEigenvectorNoisy2021}%
  \BibitemOpen
  \bibfield  {author} {\bibinfo {author} {\bibfnamefont {B.}~\bibnamefont
  {Koczor}},\ }\bibfield  {title} {\bibinfo {title} {The {{Dominant
  Eigenvector}} of a {{Noisy Quantum State}}},\ }\href
  {http://arxiv.org/abs/2104.00608} {\bibfield  {journal} {\bibinfo  {journal}
  {arXiv:2104.00608 [quant-ph]}\ } (\bibinfo {year}
  {2021}{\natexlab{b}})}\BibitemShut {NoStop}%
\bibitem [{\citenamefont {Bravyi}\ \emph {et~al.}(2017)\citenamefont {Bravyi},
  \citenamefont {Gambetta}, \citenamefont {Mezzacapo},\ and\ \citenamefont
  {Temme}}]{bravyiTaperingQubitsSimulate2017}%
  \BibitemOpen
  \bibfield  {author} {\bibinfo {author} {\bibfnamefont {S.}~\bibnamefont
  {Bravyi}}, \bibinfo {author} {\bibfnamefont {J.~M.}\ \bibnamefont
  {Gambetta}}, \bibinfo {author} {\bibfnamefont {A.}~\bibnamefont
  {Mezzacapo}},\ and\ \bibinfo {author} {\bibfnamefont {K.}~\bibnamefont
  {Temme}},\ }\bibfield  {title} {\bibinfo {title} {Tapering off qubits to
  simulate fermionic {{Hamiltonians}}},\ }\href
  {http://arxiv.org/abs/1701.08213} {\bibfield  {journal} {\bibinfo  {journal}
  {arXiv:1701.08213 [quant-ph]}\ } (\bibinfo {year} {2017})}\BibitemShut
  {NoStop}%
\bibitem [{\citenamefont {Jiang}\ \emph {et~al.}(2019)\citenamefont {Jiang},
  \citenamefont {McClean}, \citenamefont {Babbush},\ and\ \citenamefont
  {Neven}}]{jiangMajoranaLoopStabilizer2019}%
  \BibitemOpen
  \bibfield  {author} {\bibinfo {author} {\bibfnamefont {Z.}~\bibnamefont
  {Jiang}}, \bibinfo {author} {\bibfnamefont {J.}~\bibnamefont {McClean}},
  \bibinfo {author} {\bibfnamefont {R.}~\bibnamefont {Babbush}},\ and\ \bibinfo
  {author} {\bibfnamefont {H.}~\bibnamefont {Neven}},\ }\bibfield  {title}
  {\bibinfo {title} {Majorana {{Loop Stabilizer Codes}} for {{Error
  Mitigation}} in {{Fermionic Quantum Simulations}}},\ }\href
  {https://doi.org/10.1103/PhysRevApplied.12.064041} {\bibfield  {journal}
  {\bibinfo  {journal} {Physical Review Applied}\ }\textbf {\bibinfo {volume}
  {12}},\ \bibinfo {pages} {064041} (\bibinfo {year} {2019})}\BibitemShut
  {NoStop}%
\bibitem [{\citenamefont {Derby}\ \emph {et~al.}(2021)\citenamefont {Derby},
  \citenamefont {Klassen}, \citenamefont {Bausch},\ and\ \citenamefont
  {Cubitt}}]{derbyCompactFermionQubit2021a}%
  \BibitemOpen
  \bibfield  {author} {\bibinfo {author} {\bibfnamefont {C.}~\bibnamefont
  {Derby}}, \bibinfo {author} {\bibfnamefont {J.}~\bibnamefont {Klassen}},
  \bibinfo {author} {\bibfnamefont {J.}~\bibnamefont {Bausch}},\ and\ \bibinfo
  {author} {\bibfnamefont {T.}~\bibnamefont {Cubitt}},\ }\bibfield  {title}
  {\bibinfo {title} {Compact fermion to qubit mappings},\ }\href
  {https://doi.org/10.1103/PhysRevB.104.035118} {\bibfield  {journal} {\bibinfo
   {journal} {Physical Review B}\ }\textbf {\bibinfo {volume} {104}},\ \bibinfo
  {pages} {035118} (\bibinfo {year} {2021})}\BibitemShut {NoStop}%
\bibitem [{\citenamefont {O'Brien}\ \emph {et~al.}(2021)\citenamefont
  {O'Brien}, \citenamefont {Polla}, \citenamefont {Rubin}, \citenamefont
  {Huggins}, \citenamefont {McArdle}, \citenamefont {Boixo}, \citenamefont
  {McClean},\ and\ \citenamefont
  {Babbush}}]{obrienErrorMitigationVerified2021}%
  \BibitemOpen
  \bibfield  {author} {\bibinfo {author} {\bibfnamefont {T.~E.}\ \bibnamefont
  {O'Brien}}, \bibinfo {author} {\bibfnamefont {S.}~\bibnamefont {Polla}},
  \bibinfo {author} {\bibfnamefont {N.~C.}\ \bibnamefont {Rubin}}, \bibinfo
  {author} {\bibfnamefont {W.~J.}\ \bibnamefont {Huggins}}, \bibinfo {author}
  {\bibfnamefont {S.}~\bibnamefont {McArdle}}, \bibinfo {author} {\bibfnamefont
  {S.}~\bibnamefont {Boixo}}, \bibinfo {author} {\bibfnamefont {J.~R.}\
  \bibnamefont {McClean}},\ and\ \bibinfo {author} {\bibfnamefont
  {R.}~\bibnamefont {Babbush}},\ }\bibfield  {title} {\bibinfo {title} {Error
  {{Mitigation}} via {{Verified Phase Estimation}}},\ }\href
  {https://doi.org/10.1103/PRXQuantum.2.020317} {\bibfield  {journal} {\bibinfo
   {journal} {PRX Quantum}\ }\textbf {\bibinfo {volume} {2}},\ \bibinfo {pages}
  {020317} (\bibinfo {year} {2021})}\BibitemShut {NoStop}%
\bibitem [{\citenamefont {Eisert}\ \emph {et~al.}(2020)\citenamefont {Eisert},
  \citenamefont {Hangleiter}, \citenamefont {Walk}, \citenamefont {Roth},
  \citenamefont {Markham}, \citenamefont {Parekh}, \citenamefont {Chabaud},\
  and\ \citenamefont {Kashefi}}]{eisertQuantumCertificationBenchmarking2020}%
  \BibitemOpen
  \bibfield  {author} {\bibinfo {author} {\bibfnamefont {J.}~\bibnamefont
  {Eisert}}, \bibinfo {author} {\bibfnamefont {D.}~\bibnamefont {Hangleiter}},
  \bibinfo {author} {\bibfnamefont {N.}~\bibnamefont {Walk}}, \bibinfo {author}
  {\bibfnamefont {I.}~\bibnamefont {Roth}}, \bibinfo {author} {\bibfnamefont
  {D.}~\bibnamefont {Markham}}, \bibinfo {author} {\bibfnamefont
  {R.}~\bibnamefont {Parekh}}, \bibinfo {author} {\bibfnamefont
  {U.}~\bibnamefont {Chabaud}},\ and\ \bibinfo {author} {\bibfnamefont
  {E.}~\bibnamefont {Kashefi}},\ }\bibfield  {title} {\bibinfo {title} {Quantum
  certification and benchmarking},\ }\href
  {https://doi.org/10.1038/s42254-020-0186-4} {\bibfield  {journal} {\bibinfo
  {journal} {Nature Reviews Physics}\ }\textbf {\bibinfo {volume} {2}},\
  \bibinfo {pages} {382} (\bibinfo {year} {2020})}\BibitemShut {NoStop}%
\bibitem [{\citenamefont {Strikis}\ \emph {et~al.}(2021)\citenamefont
  {Strikis}, \citenamefont {Qin}, \citenamefont {Chen}, \citenamefont
  {Benjamin},\ and\ \citenamefont {Li}}]{strikisLearningbasedQuantumError2021}%
  \BibitemOpen
  \bibfield  {author} {\bibinfo {author} {\bibfnamefont {A.}~\bibnamefont
  {Strikis}}, \bibinfo {author} {\bibfnamefont {D.}~\bibnamefont {Qin}},
  \bibinfo {author} {\bibfnamefont {Y.}~\bibnamefont {Chen}}, \bibinfo {author}
  {\bibfnamefont {S.~C.}\ \bibnamefont {Benjamin}},\ and\ \bibinfo {author}
  {\bibfnamefont {Y.}~\bibnamefont {Li}},\ }\bibfield  {title} {\bibinfo
  {title} {Learning-based quantum error mitigation},\ }\href
  {http://arxiv.org/abs/2005.07601} {\bibfield  {journal} {\bibinfo  {journal}
  {arXiv:2005.07601 [quant-ph]}\ } (\bibinfo {year} {2021})}\BibitemShut
  {NoStop}%
\bibitem [{\citenamefont {Czarnik}\ \emph {et~al.}(2020)\citenamefont
  {Czarnik}, \citenamefont {Arrasmith}, \citenamefont {Coles},\ and\
  \citenamefont {Cincio}}]{czarnikErrorMitigationClifford2020}%
  \BibitemOpen
  \bibfield  {author} {\bibinfo {author} {\bibfnamefont {P.}~\bibnamefont
  {Czarnik}}, \bibinfo {author} {\bibfnamefont {A.}~\bibnamefont {Arrasmith}},
  \bibinfo {author} {\bibfnamefont {P.~J.}\ \bibnamefont {Coles}},\ and\
  \bibinfo {author} {\bibfnamefont {L.}~\bibnamefont {Cincio}},\ }\bibfield
  {title} {\bibinfo {title} {Error mitigation with {{Clifford}} quantum-circuit
  data},\ }\href {http://arxiv.org/abs/2005.10189} {\bibfield  {journal}
  {\bibinfo  {journal} {arXiv:2005.10189 [quant-ph]}\ } (\bibinfo {year}
  {2020})}\BibitemShut {NoStop}%
\bibitem [{\citenamefont {Montanaro}\ and\ \citenamefont
  {Stanisic}(2021)}]{montanaroErrorMitigationTraining2021}%
  \BibitemOpen
  \bibfield  {author} {\bibinfo {author} {\bibfnamefont {A.}~\bibnamefont
  {Montanaro}}\ and\ \bibinfo {author} {\bibfnamefont {S.}~\bibnamefont
  {Stanisic}},\ }\bibfield  {title} {\bibinfo {title} {Error mitigation by
  training with fermionic linear optics},\ }\href
  {http://arxiv.org/abs/2102.02120} {\bibfield  {journal} {\bibinfo  {journal}
  {arXiv:2102.02120 [quant-ph]}\ } (\bibinfo {year} {2021})}\BibitemShut
  {NoStop}%
\bibitem [{\citenamefont {McClean}\ \emph {et~al.}(2020)\citenamefont
  {McClean}, \citenamefont {Jiang}, \citenamefont {Rubin}, \citenamefont
  {Babbush},\ and\ \citenamefont {Neven}}]{mccleanDecodingQuantumErrors2020}%
  \BibitemOpen
  \bibfield  {author} {\bibinfo {author} {\bibfnamefont {J.~R.}\ \bibnamefont
  {McClean}}, \bibinfo {author} {\bibfnamefont {Z.}~\bibnamefont {Jiang}},
  \bibinfo {author} {\bibfnamefont {N.~C.}\ \bibnamefont {Rubin}}, \bibinfo
  {author} {\bibfnamefont {R.}~\bibnamefont {Babbush}},\ and\ \bibinfo {author}
  {\bibfnamefont {H.}~\bibnamefont {Neven}},\ }\bibfield  {title} {\bibinfo
  {title} {Decoding quantum errors with subspace expansions},\ }\href
  {https://doi.org/10.1038/s41467-020-14341-w} {\bibfield  {journal} {\bibinfo
  {journal} {Nature Communications}\ }\textbf {\bibinfo {volume} {11}},\
  \bibinfo {pages} {636} (\bibinfo {year} {2020})}\BibitemShut {NoStop}%
\end{thebibliography}

\end{document}